\documentclass[aps,prx,twocolumn,nofootinbib,longbibliography,notitlepage]{revtex4-1}
\usepackage{times}
\usepackage{graphicx}
\usepackage{tabularx}
\usepackage{amsmath}
\usepackage{amstext}
\usepackage{amssymb}
\usepackage{xfrac}
\usepackage[colorlinks,citecolor=blue]{hyperref}
\usepackage{graphicx}
\usepackage{amsmath}
\usepackage{amstext}
\usepackage{amssymb}
\usepackage{amsfonts}
\usepackage{longtable,booktabs}
\usepackage{hyperref}
\usepackage{url}
\usepackage{subfigure}
\usepackage{dsfont}
\usepackage{booktabs}
\usepackage{amsbsy}
\usepackage{dcolumn}
\usepackage{amsthm}

\usepackage{bm}
\usepackage{esint}
\usepackage{multirow}
\usepackage{hyperref}
\usepackage{cleveref}\usepackage{amsmath}
\usepackage{amsfonts}
\usepackage{amssymb}
\usepackage{amsfonts}
\usepackage{amsbsy}
\usepackage{dcolumn}
\usepackage{bm}
\usepackage{multirow}
\usepackage{color}
\usepackage{extarrows}
\usepackage{datetime}
\usepackage{comment}
\usepackage[super]{nth}
\usepackage{tikz-cd}

\hypersetup{
    colorlinks=magenta,
    linkcolor=blue,
    filecolor=magenta,
        urlcolor=magenta,
}
\def\Z{\mathbb{Z}}

\newcommand{\bra}{\langle}
\newcommand{\ket}{\rangle}

\newcommand{\LRE}{\mathtt{LRE}}
\newcommand{\ERG}{\mathtt{ERG}}
\newcommand{\CNOT}{\mathtt{CNOT}}
\newcommand{\Otimes}{\bm{\pmb{\,\otimes\,}}}

\def\Z{\mathbb{Z}}

\usepackage{extarrows}
\usepackage{datetime}

\usepackage{comment}

\usepackage{lineno}

\begin{document}
	
	%\preprint{}
	
	\title{Hierarchy of Entanglement Renormalization and Long-Range Entangled States}
 	
	
	
	\author{Meng-Yuan Li}
	\affiliation{School of Physics, State Key Laboratory of Optoelectronic Materials and Technologies, and Guangdong Provincial Key Laboratory of Magnetoelectric Physics and Devices, Sun Yat-sen University, Guangzhou, 510275, China}
	\author{Peng Ye}
	\email{yepeng5@mail.sysu.edu.cn}
	\affiliation{School of Physics, State Key Laboratory of Optoelectronic Materials and Technologies, and Guangdong Provincial Key Laboratory of Magnetoelectric Physics and Devices, Sun Yat-sen University, Guangzhou, 510275, China}
	
	%\thanks{}

 	\date{\today}

\begin{abstract}
As a quantum-informative window into quantum many-body  physics, the concept and application of entanglement renormalization group (ERG)  have been playing a vital role in the study of novel quantum phases of matter, especially long-range entangled (LRE) states in topologically ordered systems.  For instance, by recursively applying  local unitaries as well as adding/removing qubits that form  product states, the 2D toric code ground states, i.e., fixed point of $\mathbb{Z}_2$ topological order,    are efficiently coarse-grained with respect to the system size. As a further improvement, the addition/removal of 2D toric codes into/from     the ground states of the 3D X-cube   model, is shown to be indispensable and remarkably leads to well-defined fixed points of a large class of fracton orders that are non-liquid-like.   Here, we present   a  substantially  unified   ERG  framework in which     general degrees of freedom are allowed to be recursively added/removed.  Specifically, we establish  an exotic   hierarchy of ERG and  LRE states in Pauli stabilizer codes, where   the 2D toric code and 3D X-cube models are naturally included. In the hierarchy, LRE states like 3D X-cube and 3D toric code  ground states can be added/removed in ERG processes of more complex LRE states. In this way, a large group of Pauli stabilizer codes  are categorized into  a series of ``state towers'';   with each tower, in addition to local unitaries including  $\mathtt{CNOT}$ (controlled-NOT) gates, lower LRE states of level-$n$ are   added/removed in the level-$n$ ERG process of an upper LRE state of level-$(n+1)$, connecting LRE states of different levels  and unveiling   complex relations among LRE states.   As future directions, we expect this hierarchy can be applied to more general LRE states, leading to a unified ERG scenario of LRE states and exact tensor-network representations in the form of more generalized branching MERA (Multiscale Entanglement Renormalization Ansatz).

 \end{abstract}

%\keywords{}


\maketitle

%\tableofcontents
% References should be done using the  \cite, \ref, and \label commands

\section{Introduction}
\label{sec:intro}

For the past decades,  the goal of classification and characterization of  novel quantum phases of matter has been  indispensably intertwined with the surprisingly rapid progress on many-body  quantum entanglement ~\cite{White1992,White1993,Schollwoeck2005,Vidal2007,Aguado2008,Vidal2008,Koenig2009,Evenbly2014,Chen2010,Levin2006,Li2008,Kitaev2006b,Pollmann2010,Zeng2019,Gu2009,Wen2017}.  This line of efforts significantly reshapes  modern many-body physics from the emphasis of   entanglement structure instead of local correlation functions and local order parameters. For instance, the topologically ordered ground states of, e.g., fractional quantum Hall liquids~\cite{girvin2005introduction}, chiral spin liquids~\cite{Wen1989}, the toric code~\cite{Kitaev2006,Kitaev2009} and   string-net models~\cite{Levin2005a}  have been identified as long-range entangled (LRE)   states~\cite{Chen2010} that cannot be adiabatically connected to (unentangled) product states by local unitary (LU) transformations, i.e., disentanglers. In contrast, short-range entangled states (SRE) can always be connected to  product states by LU transformations. In particular, symmetry protected topological states (SPT)~\cite{Chen2012}, e.g., the Haldane spin chain, are a special class of SRE states in which all above-mentioned LU transformations inevitably  break the global symmetry that protects SPT order. Remarkably, a series of stabilizer code models realizing topological orders are found to be fixed points of certain entanglement renormalization group (ERG) transformations~\cite{Vidal2007,Aguado2008,Koenig2009} that simultaneously lead  to an efficient representation of the topologically ordered ground state in terms of a tensor network, the multiscale entanglement renormalization ansatz (MERA)~\cite{Vidal2008,Koenig2009,Evenbly2014}. The idea of ERG  provides a remarkable quantum-informative framework that significantly revolutionizes the traditional real-space and momentum-space  renormalization-group treatments of quantum many-body systems and quantum field theory. More specifically, during the process of ERG transformations,     LU transformations and addition/removal of product states are recursively  performed, such that the number of qubits (i.e., the system size) and short-range entanglement can be coarse-grained while the long-range entanglement patterns (e.g., braiding and fusion data of 2D anyon systems) keep unaltered. 

Recently, the concept of ERG transformations has been substantially advanced in order to unveil the quantum entanglement structure and   fixed points of  \emph{fracton orders}---an exotic class of topologically ordered non-liquids~\cite{Shirley2018,Haah2014,Swingle2016,Swingle2016a,Dua2020, Shirley2019b, Wang2019c, Shirley2022, WenXG_nonliquid2020,WangJuven_nonliquid2022, Nandkishore2019,Pretko2020}. In contrast to ``pure'' topological orders (e.g., the fractional quantum Hall states) that are liquid states, fracton orders are a kind of non-liquid-like LRE states whose local Hamiltonians support ground state degeneracy (GSD) that not only is locally indistinguishable (thus topologically ordered) but also grows subextensively with respect to the system size.   For example, the GSD of  X-cube model---the prototypical example of type-I fracton order---on a $3$-torus satisfies that $\text{log}_2 GSD$ grows linearly with the linear system size $L$~\cite{Vijay2016}. Immediately, it has been discovered that, to consistently define quantum phases and fixed points of fracton orders in the framework of entanglement renormalization, not only product states (i.e., SRE states), but also ``pure'' topological orders (i.e., a kind of LRE states) defined on  lower dimensional space should be added/removed, such that  two X-cube ground states of different system sizes can be adiabatically   connected~\cite{Shirley2018}.  
\begin{figure}
	\centering
	\includegraphics[width=1.0\linewidth]{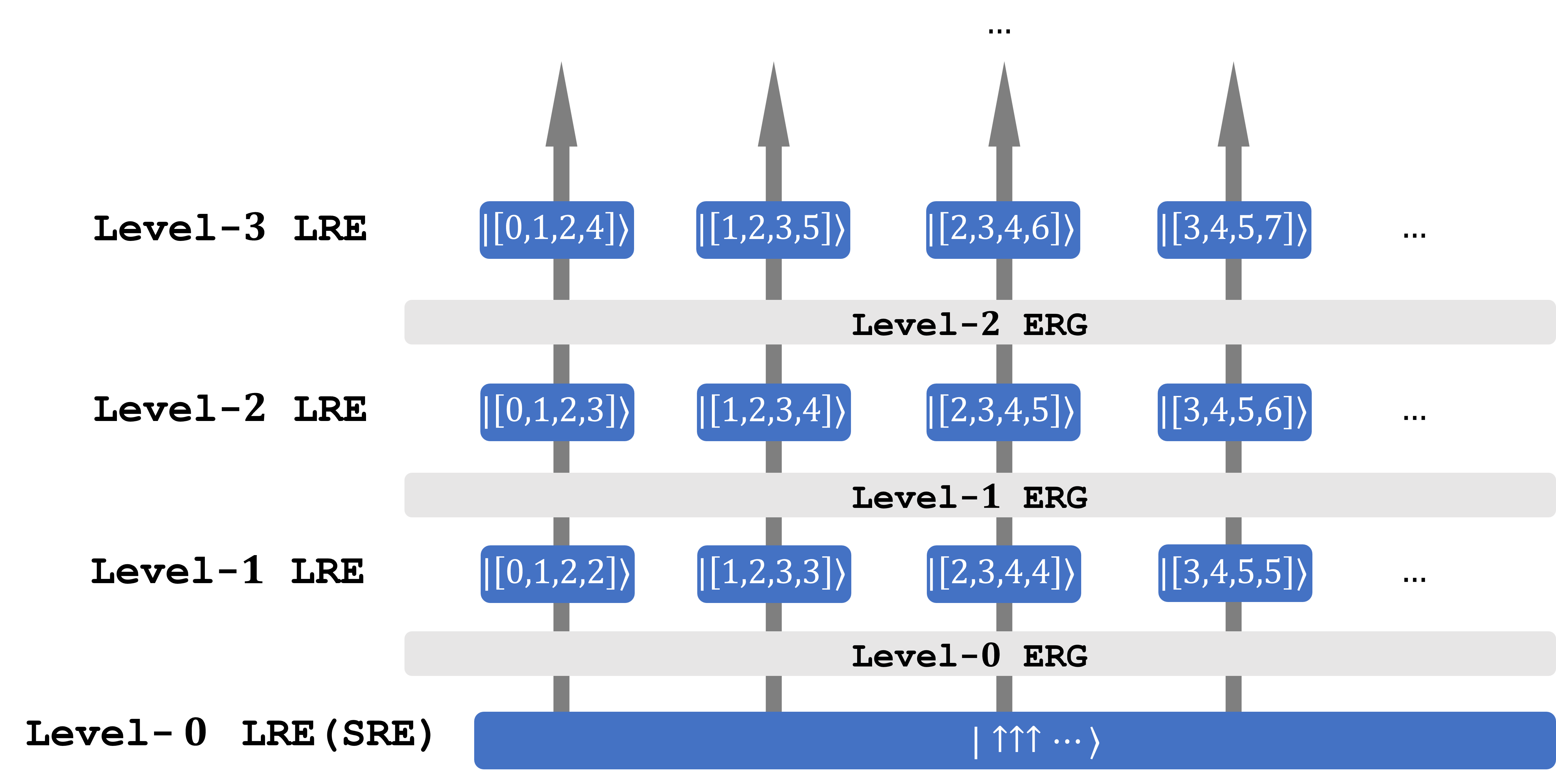}
	\caption{\textbf{Illustration of state towers that exhibit a hierarchy of entanglement renormalization and long-range entangled states}. For concreteness,  a class of Pauli stabilizer codes  are studied, each of which is labeled by four integers.     $[0,1,2,2]$, $[1,2,3,3]$ and $[0,1,2,3]$ are respectively 2D toric code, 3D toric code and 3D X-cube models. $|[\cdots]\rangle$ refers to a ground state of a certain model (i.e., a state in the stabilizer subspace, see Sec.~\ref{sec:pre}). There are  a  series of  {state towers} denoted by upward arrows, and along each arrow, a lower $\mathtt{LRE}^n$ state can be added/removed in the $\mathtt{ERG}^n$ process of the upper $\mathtt{LRE}^{n+1}$ state as demonstrated in Eq.~(\ref{eq_ERG_relation}). To unify the notation, SRE states (including unentangled product states) are symbolically denoted as $\mathtt{LRE}^0$.}
	\label{fig:tree}
\end{figure} 
Despite the success of such ERG generalization, whether or not there is a much deeper mechanism towards a unified ERG framework is yet to be investigated.

In this paper, through exactly solvable models, we present a unified ERG framework via a hierarchical structure of ERG as well as the associated LRE states,  where the above ERG transformations of original definition~\cite{Vidal2007,Aguado2008,Koenig2009} and that of X-cube model~\cite{Shirley2018} are naturally included.  To be more specific, as shown in Fig.~\ref{fig:tree}, we construct ERG transformations for a series of Pauli stabilizer code models~\cite{Gottesman1997} proposed in Ref.~\cite{Li2020}, such that the models are fixed points of ERG transformations.  All models we will study in this paper are uniquely denoted by  four integers, i.e., $[d_n,d_s,d_l,D]$, where  a subset labeled by $[d,d+1,d+2,D]$ is found to be Pauli stabilizer code models with emergent $\Z_2$ gauge symmetry (see Sec.~\ref{sec:pre} for more details). The familiar 3D X-cube model is denoted as $[0,1,2,3]$. And, we also successfully incorporate toric code models of all dimensions into the labeling system, which has not be included in Ref.~\cite{Li2020}. For example, the 2D toric code model is labeled by $[0,1,2,2]$.   Remarkably, in the ERG transformations of these Pauli stabilizer codes, we find a  hierarchical structure summarized in Fig.~\ref{fig:tree}: in an ERG transformation connecting two $[d,d+1,d+2,D]$ states (i.e., the ground states of $[d,d+1,d+2,D]$ model as lattice Hamiltonian) with $D > d + 2$ of different sizes, $[d,d+1,d+2,D-1]$ states are added/removed in addition to local unitaries (e.g., $\mathtt{CNOT}$), such that all Pauli stabilizer codes are fixed-points of the ERG transformations. While the $\log_2 GSD$ of these topological non-liquid models grows polynomially  with respect to the linear system size~\cite{Li2021}, such ERG transformations are found to keep the GSD formulas consistent in different length scales. All in all, the ERG relation can be symbolically expressed as follows:
 $ 	|[d,d+1,d+2,D]\ket \sim |[d,d+1,d+2,D]\ket' \Otimes |[d,d+1,d+2,D-1]\ket,
$ where $|[d,d+1,d+2,D]\ket$ and $|[d,d+1,d+2,D]\ket'$ are $[d,d+1,d+2,D]$ states of different sizes, and $\sim$ means the two sides can be connected by an LU transformation. 

In the unified framework, ERG transformations obey the following   rules:
\begin{itemize}
	\item[$-$] In the   ERG transformations on Pauli stabilizer codes considered here, LRE states are categorized into different levels, denoted as $\mathtt{LRE}^n$ with  the level index $n=0,1,2,\cdots$. Unentangled product states and more general SRE states are dubbed  ``level-$0$ LRE states''  (denoted as $\mathtt{LRE}^0$ symbolically) for the notational convenience; 
	\item[$-$] ERG transformations where level-$n$ LRE states are added/removed are dubbed ``{level-$n$ ERG}'' (denoted as $\mathtt{ERG}^n$ symbolically) transformations. Unless otherwise specified, $n$ is the highest level of added/removed LRE states;
	\item[$-$] States of the same stabilizer code with different sizes that can be connected by level-$n$ ERG transformations are identified as  $\mathtt{LRE}^{n+1}$.
\end{itemize} 
Then an $\mathtt{ERG}^n$ transformation can be symbolically expressed as follows:
\begin{align}
 \mathtt{ERG}^n: \,\,\mathtt{LRE}^{n+1} \sim \mathtt{LRE}^{n+1}\Otimes \mathtt{LRE}^n  \label{eq_ERG_relation}
\end{align}
which explicitly shows a hierarchy of ERG transformations as well as LRE states along each upward arrow in Fig.~\ref{fig:tree}.   For example, the ERG of  the 2D toric code is given by $\mathtt{ERG}^{0}$~\cite{Vidal2007,Aguado2008,Koenig2009,Zeng2019}:
 \begin{align}
 \mathtt{ERG}^0:\,\,	\mathtt{LRE}^{1} \sim \mathtt{LRE}^{1} \Otimes \mathtt{LRE}^0\,,  \label{eq_ERG_relation_ToricCode}
\end{align}
where a toric code ground state is denoted as $\mathtt{LRE}^{1}$ and product states denoted as $\mathtt{LRE}^0$ are added/removed (note that SRE states are also symbolically denoted as $\LRE^0$ for the notational convenience). Similarly, the ERG of the 3D X-cube model is given by $\mathtt{ERG}^1$~\cite{Shirley2018}:  
\begin{align}
 \mathtt{ERG}^1:\,\,	\mathtt{LRE}^{2} \sim \mathtt{LRE}^{2} \Otimes \mathtt{LRE}^1 \,, \label{eq_ERG_relation_Xcube}
\end{align}
where an X-cube ground state is denoted as $\mathtt{LRE}^{2}$ and the 2D toric code ground state $\mathtt{LRE}^1$  is added/removed. 

We also note that, the above rules are established in the concrete stabilizer codes studied in this paper.  In fact, Eq.~(\ref{eq_ERG_relation}) may be in principle a concrete realization of  the following more general level-$m$ ERG transformation denoted as $\mathtt{ERG}^m$:
 \begin{align}
 \mathtt{ERG}^m:\,\,	\mathtt{LRE}^{n} \sim \mathtt{LRE}^{n} \Otimes \mathtt{LRE}^m \,, \label{eq_ERG_relation_general}
\end{align}
where $m<n$ is generally required. Assuming the existence of such $\ERG^m$ transformations, a natural conjecture is that the level of LRE states may be decided by the $\ERG^m$ transformations of the highest possible level. We leave such general ERG transformations as well as implied MERAs to further exploration.

The reminder of this paper is organized as follows.   In Sec.~\ref{sec:pre}, we introduce some very useful geometric notations used in this paper and give a brief introduction to the $[d,d+1,d+2,D]$ models that  include the 2D  toric code model and  the 3D X-cube model as special examples. Especially, we explain how to incorporate toric codes into the labeling system. Sec.~\ref{sec:hier} is dedicated to a detailed demonstration of some concrete ERG transformations. In Sec.~\ref{subsec:glu}, as a warm-up, we perform  the ERG transformations on  the 2D toric code model (denoted as $[0,1,2,2]$), while an alternative approach was reviewed in Appendix~\ref{appendix_toriccode} by following Ref.~\cite{Zeng2019}.  In Sec.~\ref{subsec:ffo}, we   review the  ERG transformations on the 3D X-cube model (denoted as $[0,1,2,3]$).  Then, we concretely construct the ERG transformations of different levels for $[0,1,2,4]$ and $[1,2,3,4]$ models respectively in Sec.~\ref{subsec:0124_ER} and \ref{subsec:1234_ER}. Sec.~\ref{sec:general} is dedicated to  ERG transformations in general $[d,d+1,d+2,D]$ models. In Sec.~\ref{subsec:dddD_ER}, we demonstrate a general recipe for the ERG transformations of general $[d,d+1,d+2,D]$ models.  In Sec.~\ref{subsec:proof_ham}, we prove that the models are indeed fixed points of corresponding ERG transformations. Then, we demonstrate how these ERG transformations lead to the concept of a hierarchy of ERG transformations and LRE states in Sec.~\ref{subsec:dis}. A summary and outlook  is given in  Sec.~\ref{sec:outlook}.

\section{Labeling system of Pauli stabilizer codes}
\label{sec:pre}
This section is dedicated to the introduction of some background, including geometric notations and a family of Pauli stabilizer code models denoted by $[d,d+1,d+2,D]$. Specially, we notice that $[D-2,D-1,D,D]$ models can be regarded as a $D$-dimensional generalization of the 2D toric code model.  

\subsection{Geometric notations and lattice Hamiltonians}

In this paper we need to involve some discussion about high dimensional geometric objects, so we believe it is beneficial to at first introduce some relevant notations. For a hypercubic lattice discussed in this paper, unless otherwise specified, we set lattice constant to be $1$. Then, we introduce the concept of $n$-cubes denoted by $\gamma_n$, that simply refers to $n$-dimensional analogs of cube. For example, a $\gamma_0$ ($0$-cube) is simply a vertex, a $\gamma_1$ ($1$-cube) is a link, a $\gamma_2$ ($2$-cube) is a plaquette and a $\gamma_3$ ($3$-cube) is a conventional cube. In a $D$-dimensional hypercubic lattice, with the above notations, we can use the coordinates of the center of a $\gamma_n$ ($n<D$ is assumed) to refer to the $\gamma_n$ itself, as such a $\gamma_n$ can be uniquely determined by the coordinates. Besides, we can see that the coordinate representation of a $\gamma_n$ in a $D$-dimensional hypercubic lattice is always composed of $n$ half-odd-integers (or hald-integer in shorthand) and $(D-n)$ integers. For example, in 3D cubic lattice, the coordinate representation of a $\gamma_2$ (i.e. plaquette), such as $(\frac{1}{2}, \frac{1}{2}, 0)$ and $(\frac{7}{2}, 5, \frac{1}{2})$, always contains \textit{two} half-integers and \textit{one} integer. What's more, following the terminology in Ref.~\cite{Li2020}, we say a $n$-cube $\gamma_{n} = (x_1,x_2,\cdots,x_D)$ and an $m$-cube $\gamma_{m}=(y_1,y_2,\cdots,y_D)$ to be nearest to each other when $|x_1-y_1|+|x_2-y_2|+\cdots+|x_D-y_D|=\frac{|m-n|}{2}$ for $m\neq n$. Specially, when $m=n$, we say they are nearest to each when $|x_1-y_1|+|x_2-y_2|+\cdots+|x_D-y_D|=1$. We can check that such a definition of being nearest is consistent with the usual conventions.

Next, we give a brief review of the definition of $[d,d+1,d+2,D]$ Pauli stabilizer code models. As lattice Hamiltonians, $[d,d+1,d+2,D]$ models is a subset of $[d_n,d_s,d_l,D]$ models proposed in Ref.~\cite{Li2020}. In general, a $[d_n,d_s,d_l,D]$ model is defined on a $D$-dimensional hypercubic lattice, with one $\frac{1}{2}$-spin defined on each $d_s$-cube (i.e. $\gamma_{d_s}$). And the Hamiltonian is given as follows:
\begin{align}
	\label{eq:ham_nslD}
	H_{[d_n,d_s,d_l,D]} = -\sum_{\gamma_D} A_{\gamma_D}- \sum_{\gamma_{d_n}} \sum_l B^l_{\gamma_{d_n}},
\end{align}
where a $B^l_{\gamma_{d_n}}$ term is the product of the $z$-components of the spins (a) being nearest to the $d_n$-cube $\gamma_{d_n}$ and (b) living in a $d_l$-dimensional subsystem given by index $l$, and an $A_{\gamma_D}$ term is the product of the $x$-components of the spins being nearest to the $D$-cube $\gamma_D$. Here for simplicity, all coefficients of terms have been set to be $-1$. A concrete example of the Hamiltonian of $[0,1,2,3]$ (a.k.a. 3D X-cube) model is illustrated in Fig.~\ref{fig:models}(b). In Ref.~\cite{Li2020}, $d_n<d_s<d_l<D$ is assumed, while in this paper, we allow the case $d_l=D$ to give a more complete picture of the hierarchy of ERG transformations and LRE states. More details of this case are given in Sec.~\ref{subsec:-2-100}.

\subsection{Incorporating toric codes}
\label{subsec:-2-100}

In this paper we primarily focus on $[d,d+1,d+2,D]$ models (i.e.,   we set $d_n=d$, $d_s=d+1$, $d_l=d+2$). Here, we notice that 2D and 3D toric code models can also be included into the above model series as $[0,1,2,2]$ and $[1,2,3,3]$ models respectively. In fact, generally a $[D-2,D-1,D,D]$ model can be recognized as a $D$-dimensional generalization of 2D toric code model. Here, because $[D-2,D-1,D,D]$ models do not satisfy the $d_l<D$ condition, now the superscripts of $B$ terms are redundant, and a $B_{\gamma_{D-2}}$ term is simply the product of the $z$-components of the $4$ spins nearest to the $\gamma_{D-2}$.

To see the equivalence between a $[D-2,D-1,D,D]$ model and a $D$-dimensional toric code model, we can consider a duality, where $\gamma_n$'s are mapped to $\gamma_{D-n}$'s, that can be concretely realized by shifting the coordinates of all $\gamma_n$'s by  $(\frac{1}{2}, \frac{1}{2},\cdots,\frac{1}{2})$.  {For a given $[D-2,D-1,D,D]$ model, upon the duality, we obtain a dual model that is still defined on a $D$-dimensional hypercubic lattice, but $\frac{1}{2}$-spins originally defined on $\gamma_{D-1}$'s are now defined on $\gamma_1$'s (a.k.a. links). As for the Hamiltonian terms, the original $A_{\gamma_D}$ terms defined on $\gamma_D$'s are mapped to $A_v$ terms defined on $\gamma_0$'s (a.k.a. vertices), and the original $B_{\gamma_{D-2}}$ terms defined on $\gamma_{D-2}$'s are mapped to $B_p$ terms defined on $\gamma_2$'s (a.k.a. plaquettes).}

\begin{figure}
	\centering
	\includegraphics[width=1\linewidth]{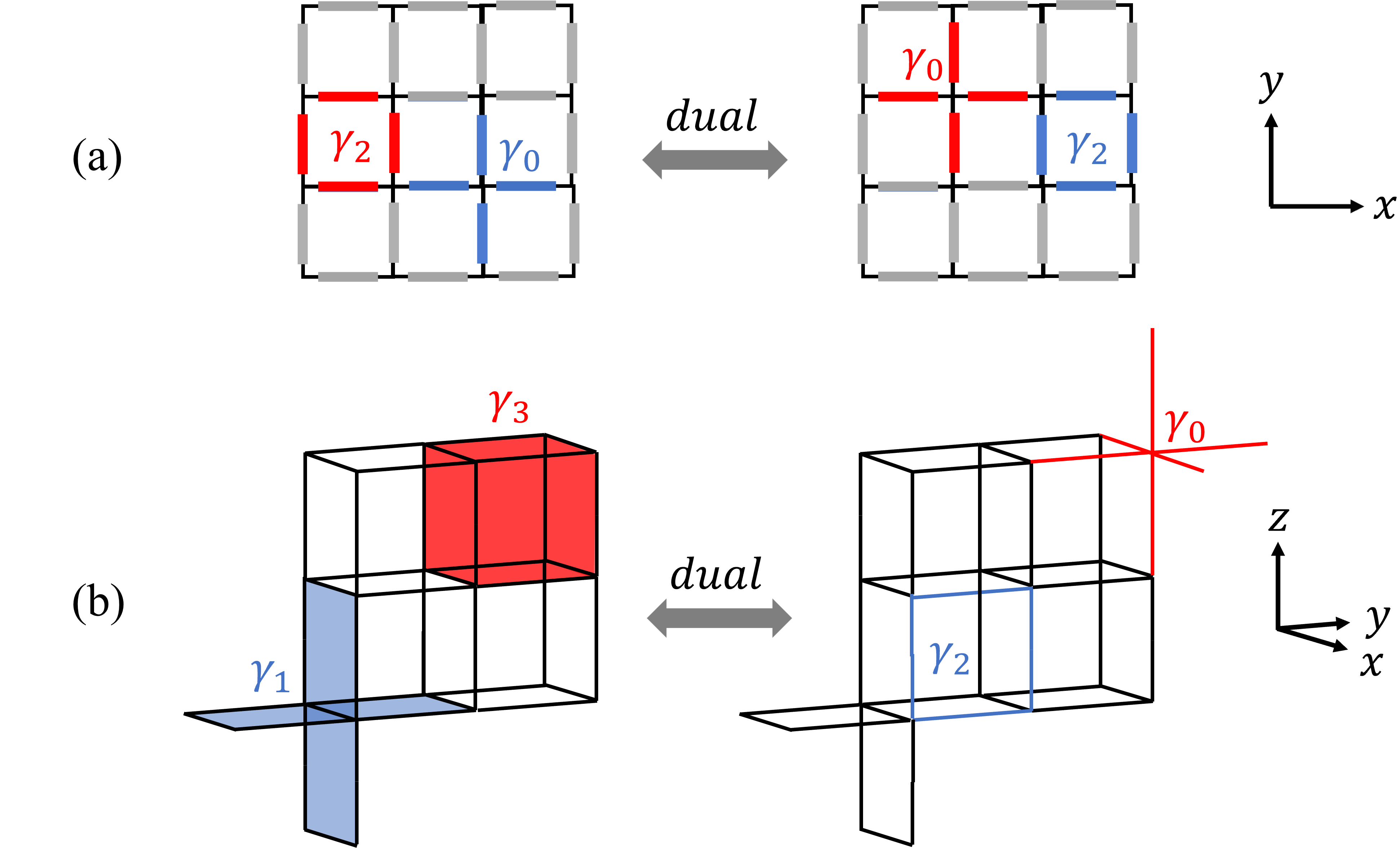}
	\caption{\textbf{Duality of $[0,1,2,2]$ and $[1,2,3,3]$ models.} Here we show two examples of the duality between original $[D-2,D-1,D,D]$ models (on left hand side) and corresponding dual models (on right hand side). (a) and (b) respectively demonstrate the duality of $[0,1,2,2]$ and $[1,2,3,3]$ models. In (a), the spins are defined on links (i.e., $\gamma_1$'s) on both sides, thus we use bars on links to refer to spins; an $A_{\gamma_2}$ Hamiltonian term highlighted with red originally defined on a plaquette (i.e., $\gamma_2$) is mapped to an $A_v$ term defined on a vertex (i.e., $\gamma_0$), and a $B_{\gamma_0}$ term highlighted with blue originally defined a vertex is mapped to a $B_p$ term defined on a plaquette. In (b), the spins are respectively defined on plaquettes in the original model and links in the dual model, thus we do not explicitly show all the spins for clarity; an $A_{\gamma_3}$ Hamiltonian term highlighted with red originally defined on a cube (i.e., $\gamma_3$) is mapped to an $A_v$ term defined on a vertex, and a $B_{\gamma_1}$ term highlighted with blue originally defined a link is mapped to a $B_p$ term defined on a plaquette.}
	\label{fig:dual}
\end{figure}

In summary, the Hamiltonian of the dual model defined on a $D$-dimensional hypercubic lattice is given by $H_{dual}=-\sum_v A_v - \sum_p B_p$, where each link is assigned with a $\frac{1}{2}$-spin, $A_v$ is the product of $x$-components of spins nearest to the vertex $v$, $B_p$ is the product of $z$-components of spins nearest to the plaquette $p$ (see Fig.~\ref{fig:dual} for the pictorial demonstration of some examples). Such a Hamiltonian is a $D$-dimensional generalization of the 2D toric code model~\cite{Hamma2005}, and the ground states of which are regarded as $\LRE^1$ states realized in different spatial dimensions (i.e., $D$). Note that the dual models themselves are not a part of $[d,d+1,d+2,D]$ models, thus in this paper the original $[D-2,D-1,D,D]$ models are more involved.  

\subsection{Ground state wavefunctions}

\begin{figure}
	\centering
	\includegraphics[width=1\linewidth]{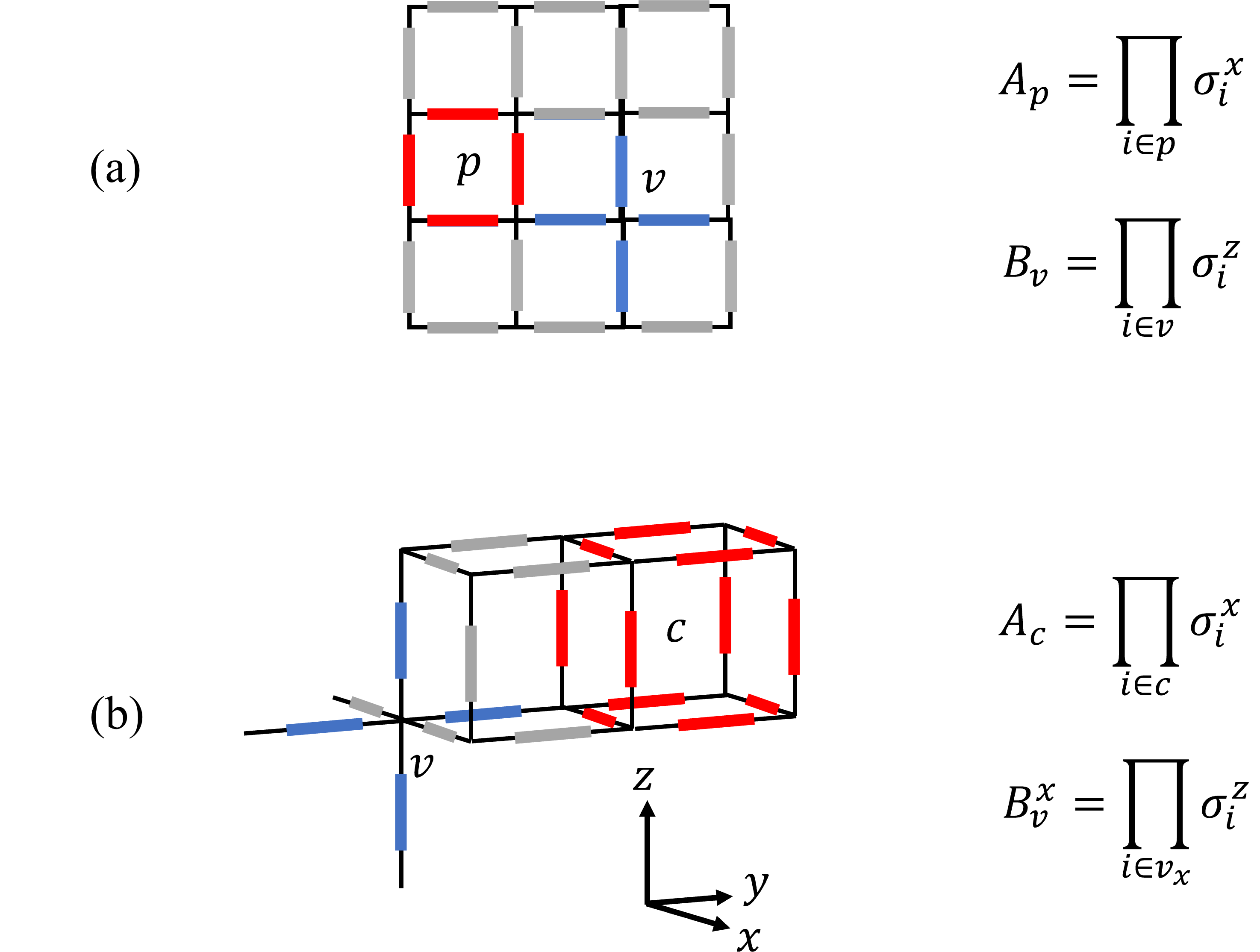}
	\caption{\textbf{Hamiltonians of some representative $[d,d+1,d+2,D]$ models.} (a) and (b) respectively demonstrate the Hamiltonian terms of the $[0,1,2,2]$ (2D toric code) and the $[0,1,2,3]$ (3D X-cube) model. In each subfigure, spins are represented by bars on links, and we draw spins acted by an $A$ term with red, and spins acted by a $B$ term with blue. In (b), we only draw a single $B^x_v$ term on vertex $v$, that is composed of the $4$ spins that are not only nearest to vertex $v$ but also located in a plane perpendicular to $x$-direction. Such $4$ spins are denoted as $i\in v_x$ in (b).}
	\label{fig:models}
\end{figure}

Then, we can use a general recipe to obtain the ground states of $[d,d+1,d+2,D]$ stabilizer code models (including $[D-2,D-1,D,D]$ models, such as 2D and 3D toric codes). The lattice Hamiltonians of these models are all of the following form:
\begin{align}
	H = -\sum_i A_i -\sum_j B_j,
\end{align}
where $i$ and $j$ are some kinds of spatial locations (e.g. vertices, centers of links and centers of plaquettes) depending on the specific model, and the index $l$ in Eq.~(\ref{eq:ham_nslD}) has been formally absorbed into index $j$ for simplicity. Here, $A_i$ and $B_j$ are respectively local products of $\sigma^x$ and $\sigma^z$ Pauli operators, and they all commute with each other (see Fig.~\ref{fig:models} for examples of $[0,1,2,2]$ and $[0,1,2,3]$ models). Therefore, a ground state $|\phi\ket$ of such a Hamiltonian has to satisfy constraints $A_i|\phi\ket = |\phi\ket,\ \forall i$ and $B_j|\phi\ket = |\phi\ket,\ \forall j$ (respectively denoted as $A$ and $B$ constraints). That is to say, for a given $[d,d+1,d+2,D]$ model, the $A_i$ and $B_j$ operators can be regarded as generators of a stabilizer group, and the ground state subspace is the corresponding stabilizer subspace~\cite{Gottesman1997, Zeng2019}, as ground states are ``stabilized'' by all $A_i$ and $B_j$ operators. In this paper, as we mainly care about the stabilizer subspaces, unless otherwise specified, for a given model, we only consider states in its ground state subspace. Then, for an arbitrary $[d,d+1,d+2,D]$ stabilizer code model, we can obtain a ground state $|\phi_n\ket$ of it by the following procedures:
\begin{itemize}
	\item[$-$] First, we consider $\sigma^z$ basis, that is to say, we use Ising configurations, where spins are denoted by their direction along $\sigma^z$, as a basis of the whole Hilbert space. For a single qubit, we use the convention $\sigma^z |\uparrow\ket = |\uparrow\ket = |0\ket = \begin{pmatrix}
	1 \\ 
	0
	\end{pmatrix} $, $\sigma^z |\downarrow\ket \!= \!-|\downarrow\ket = -|1\ket = -\begin{pmatrix}
	0 \\ 
	1
	\end{pmatrix} $ (i.e., $|0\ket$ for spin up, and $|1\ket$ for spin down).
	\item[$-$] Second, we can notice that $|0 \cdots 0 0\ket$ naturally satisfies all $B$ constraints. We denote $|0 \cdots 0 0\ket$ as the reference state.
	\item[$-$] Third, we consider the equal weight superposition of the reference state and all configurations that can be obtained by applying a series of $A_i$ operators on the reference state, and denote this state as $|\phi_n\ket$. As all $A_i$ and $B_j$ operators commute with each other, $|\phi_n\ket$ also satisfies $B$ constraints. According to our construction of $|\phi_n\ket$, where $2$ configurations that can be related by the action of $A_i$ are always equally superpositioned, we can see that $|\phi_n\ket$ must also satisfy $A$ constraints. Hence, $|\phi_n\ket$ is a ground state of the stabilizer code model.
\end{itemize} 
In the following part of this paper, we use an intuitive picture to describe an Ising configuration, by recognizing flipped spins (i.e. spin of the state $|1\ket$) as occupied by certain geometric objects. For example, if the spins are defined on links, then we recognize flipped spins as occupied by strings; and if the spins are defined on plaquettes, then we recognize flipped spins as occupied by membranes. For a $[d,d+1,d+2,D]$ model, other ground states can be obtained by applying logical operators on the $|\phi_n\ket$ state. Here in the $\sigma^z$ basis, a logical operator can be recognized as a product of a series of $\sigma^x$ operators, that commutes with all $B_j$ terms and do not equivalent to any product of a series of $A_i$ terms. For instance, in $[0,1,2,2]$ model defined on a $T^2$ ($2$-torus), such a logical operator is a non-contractible closed string composed of $\sigma^x$ operators~\cite{Kitaev2006,Kitaev2009}. 

Following this general recipe, we can see that when we ignore the topological degeneracy by focusing on the open boundary condition, we only need to consider the $|\phi_n\ket$ state, that can be regarded as a superposition of a series of configurations. For the $|\phi_n\ket$ state, $B$ terms require a superpositioned configuration to satisfy certain constraints, like flipped spins forming closed strings in $[0,1,2,2]$ model; $A$ terms require configurations that can be connected by action of $A$ terms to be equal-weight superpositioned. In Sec.~\ref{sec:hier}, a series of concrete examples are demonstrated in the corresponding subsections.

\begin{comment}
	The figure seems redundant for me now.
	\begin{figure}
		\centering
		\includegraphics[width=0.6\linewidth]{groundstates}
		\caption[$A_i$ operators of various stabilizers code models]{(a) shows an $A_p$ operator in 2D toric code models composed of $\sigma^x$ on links of a plaquette, links with $\sigma^x$ action are colored blue. (b) shows an $A_c$ operator in 3D X-cube models composed of $\sigma^x$ on links of a cage, links with $\sigma^x$ action are colored blue. (c) shows an $A_{\gamma_4}$ operator in 4D $[1,2,3,4]$ models composed of $\sigma^x$ on plaquettes around a 4D hypercube. For clarity, only $6$ plaquettes of $6$ typical directions with $\sigma^x$ action are colored blue, while in fact $A_{\gamma_4}$ is composed of $\sigma^x$ on all $24$ plaquettes of the 4D hypercube.}
		\label{fig:groundstates}
	\end{figure}
\end{comment}

\section{Hierarchy of ERG transformations and LRE states}
\label{sec:hier}

In this section, we concretely demonstrate the ERG transformations of some $[d,d+1,d+2,D]$ states. At first, in Sec.~\ref{subsec:glu} and Sec.~\ref{subsec:ffo}, we perform the ERG transformations of $[0,1,2,2]$ (2D toric code) and $[0,1,2,3]$ (3D X-cube) models respectively. Then, in Sec.~\ref{subsec:0124_ER} and Sec.~\ref{subsec:1234_ER}, we respectively construct the ERG transformations of $[0,1,2,4]$ and $[1,2,3,4]$ models.

\subsection{Level-$0$ ERG transformation of $[0,1,2,2]$ (2D toric code) states}
\label{subsec:glu}
 
The ERG transformations of $[0,1,2,2]$ states have been proposed and studied previously~\cite{Aguado2008,Koenig2009,Zeng2019}, see the review in  Appendix~\ref{appendix_toriccode}.  For consistency, here we perform an $\ERG^0$ transformation of $[0,1,2,2]$ states in an explicitly different way. This alternative  ERG process is very useful for designing    ERG transformations of other $[d,d+1,d+2,D]$ models to be discussed in this paper.

 At first, we give an intuitive picture of the $[0,1,2,2]$ states based on the general discussion in Sec.~\ref{sec:pre}. In $[0,1,2,2]$ (a.k.a. 2D toric code) model, spins are located at links of a 2D square lattice. In a superpositioned configuration of a $[0,1,2,2]$ state, a $B_j$ term requires vertex $j$ can only have $0$, $2$ or $4$ flipped nearest spins, thus flipped spins must form closed strings. An $A_i$ term flips the $4$ spins on the links of plaquette $i$, thus contractible closed strings can freely fluctuate in a ground state. We will see that, the ERG transformation indeed preserves this closed strings pattern of $[0,1,2,2]$ states. 

We start with a ground state $|\xi_i \rangle$ of  the $[0,1,2,2]$ model defined on a square lattice of the size $L_x\times L_y$ with periodic boundary condition (PBC), and obtain a ground state $|\xi_f \rangle$ on a square lattice of the size $L_x\times (L_y + 1)$ with PBC by the following transformations:

First, we choose a $T^1$ ($1$-torus, a.k.a. loop) composed of the centers of parallel links along direction $\hat{y}$ with the same $\hat{y}$-coordinate, and regard the $T^1$ as a cut: all links intersecting with the $T^1$ are cut into $2$ links. Without loss of generality, we assume the $T^1$ is located at $y=\frac{1}{2}$, which means the cut links are of the form $(i,\frac{1}{2})$, where $i$ are integers. After that, we apply a rescaling. For each cut link $l=(i,\frac{1}{2})$, we double the length of $l$ to $2$. Then, we can see that $l$ is cut into links $l_1=(i,\frac{1}{2})$ and $l_2=(i,\frac{3}{2})$ of length $1$, and now the cut $T^1$ is located at $y=1$. We assign the original spin on $l$ to $l_1$.

\begin{figure}
	\centering
	\includegraphics[width=0.9\linewidth]{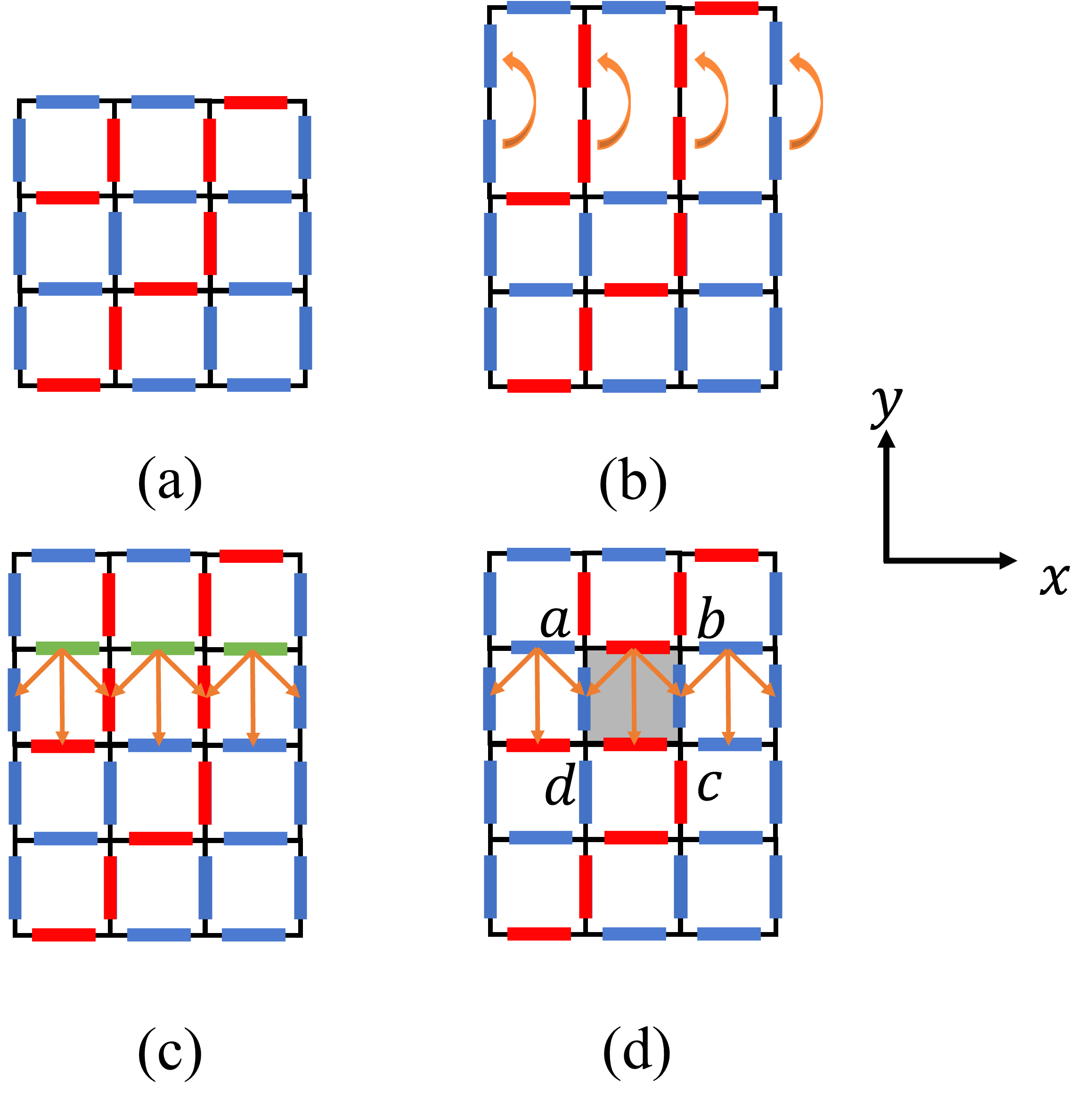}
	\caption{\textbf{ERG transformation of the 2D toric code model labeled by $[0,1,2,2]$.} This ERG transformation is denoted as $\mathtt{ERG}^0$ in Eq.~(\ref{eq_ERG_relation_ToricCode}).  In (a), a closed string configuration in the original lattice is illustrated, where $|0\ket$ spins on the links are denoted by blue bars, and $|1\ket$ spins forming strings are highlighted with red. In (b), we illustrate the string configuration after the addition of spins in state $|0\rangle$, and the added spins \textit{have been} transformed by $\mathtt{CNOT}$ gates such that the strings (formed by $|1\ket$ spins) are still closed. These $\mathtt{CNOT}$ gates targeting on the additional spins are denoted by orange arrows pointing from control qubits to target qubits. In (c) we  show the inserted $|\xi_p\rangle$ state and the $\CNOT$ gates applied on $|\xi_2\rangle$, where added spins of state $|\rightarrow \rangle$ are denoted by green bars, and $\mathtt{CNOT}$ gates are also denoted by orange arrows. In (d), we illustrate an Ising configuration of $|\xi_f\rangle$, where a concrete configuration of $|\xi_p\rangle$ is picked, and the target spins have been correspondingly transformed by the $\CNOT$ gates. We can see that, in such a configuration, flipped spins also form closed strings. Besides, an assignment of labels to the four vertices around a shadowed plaquette is also presented.      
	}
	\label{fig:toriccode_sim}
\end{figure}

Second, for each $l_2$, we put an additional spin of the state $|0\ket$ on it. It means that we enlarge the Hilbert space by taking the tensor product of the original one and the added spins, and add a series of $-\sigma_{l_2}^z$ terms to the Hamiltonian to make all the added spins in the state $|0\ket$ (since a term $-\sigma_{l_2}^z$ requires a ground state $|\phi\ket$ to satisfy $\sigma_{l_2}^z|\phi\ket=|\phi\ket$). Then, for each original cut link $l$, we apply a $\mathtt{CNOT}$ (controlled-NOT) gate with the original qubit on $l_1$ as control qubit and the added one on $l_2$ as target (see Fig.~\ref{fig:toriccode_sim} (b)). By conjugate action of $\mathtt{CNOT}$ gates, the added $-\sigma_{l_2}^z$ terms are mapped to $-\sigma_{l_1}^z \sigma_{l_2}^z$ terms (see Appendix~\ref{app_cnot}). As a result, given a cut link $l$, for an arbitrary Ising configuration $|\cdots \sigma_l \cdots \rangle$ from $|\xi_i \rangle$ ($\sigma_l = 0$ or $1$), we have $|\cdots \sigma_l \cdots \rangle \rightarrow |\cdots \sigma_{l_1} \sigma_{l_2} \cdots \rangle  $, where $\sigma_{l_1} =  \sigma_{l_2} = \sigma_{l}$. The ground state transformed by steps above is denoted as $|\xi_1\ket$.

Third, we insert a product state $|\xi_{p}\rangle = |\rightarrow \rightarrow \cdots \rightarrow \rangle$ of the size $L_x$ on the cut $T^1$ given in the first step, where $|\rightarrow \rangle=\frac{1}{\sqrt{2}} (|0\rangle +|1\rangle)$ is the eigenstate of $\sigma^x$ with eigenvalue $1$. That is to say, the spins composing the inserted state are located on links of the form $(i+\frac{1}{2}, 1)$ in the rescaled lattice (see Fig.~\ref{fig:toriccode_sim} (c), note that there are no spins on such links before this step). Then, we denote the tensor product of $|\xi_1\rangle$ and $|\xi_{p}\rangle$ as $|\xi_2\rangle = |\xi_1\rangle \Otimes |\xi_{p}\rangle$. 

Finally, we act a series of $\mathtt{CNOT}$ gates on $|\xi_2\rangle$ as illustrated in Fig.~\ref{fig:toriccode_sim} (c) and (d). The $\mathtt{CNOT}$ gates are organized in a translational invariant manner, thus we only need to specify them for a specific plaquette. Without loss of generality, we take $\gamma_2=(\frac{1}{2},\frac{1}{2})$, and denote the vertices of $\gamma_2$ by letters as shown in Fig.~\ref{fig:toriccode_sim} (d). Concretely, we have $d=(0,0), c=(1,0), a=(0,1)$ and $b=(1,1)$. Then, the $\mathtt{CNOT}$ gates can be explicitly specified as follows:
\begin{align*}
	\sigma_{ab} \rightarrow \sigma_{bc}, \sigma_{cd}, \sigma_{da},
\end{align*}
where $\sigma_{xy}$ refers to the spin located on the link between $x$ and $y$ vertices, $\rightarrow$ points from the control qubit to target qubits. 

We can straightforwardly check that after the application of the $\CNOT$ gates on $|\xi_2\rangle$, we obtain $|\xi_f\rangle$ that preserves the closed strings pattern of $[0,1,2,2]$ states. To see this, we show that the by conjugate action, the $\CNOT$ gates generate all stabilizer generators we need to obtain $[0,1,2,2]$ states on the enlarged lattice. Firstly, by regarding $|\xi_p\rangle$ as stabilized by a series of $\sigma^x$ stabilizers with $y=1$ (i.e., for each $\sigma^x$ stabilizer, the link where the stabilizer is defined satisfies $y=1$), under the conjugate action of $\CNOT$ gates, a $\sigma^x$ stabilizer with $y=1$ is mapped to an $A$ term with $y=\frac{1}{2}$; then, a $\sigma^z_{l_1} \sigma^z_{l_2}$ stabilizer obtained in the second step above is mapped to a $B$ term with $y=1$; finally, as we can notice that also in the second step above, an $A$ term with $y=\frac{1}{2}$ in the original lattice is mapped to a six-spin term composed of the $x$ components of all spins nearest to a rectangle (see Fig.~\ref{fig:toriccode_sim} (b)), by taking the product of such a modified $A$ term and an $A$ term with $y=\frac{1}{2}$, an arbitrary $A$ term with $y=\frac{3}{2}$ can be obtained. Therefore, the $|\xi_f\rangle$ is indeed a $[0,1,2,2]$ state on the enlarged lattice. Or from another perspective, the $[0,1,2,2]$ model is a fixed point of the $\ERG^0$ transformation, as symbolically expressed in Eq.~(\ref{eq_ERG_relation_ToricCode}).

\subsection{Level-$1$ ERG transformation of $[0,1,2,3]$ (3D X-cube) states}
\label{subsec:ffo}

\begin{figure}
	\centering
	\includegraphics[width=0.9\linewidth]{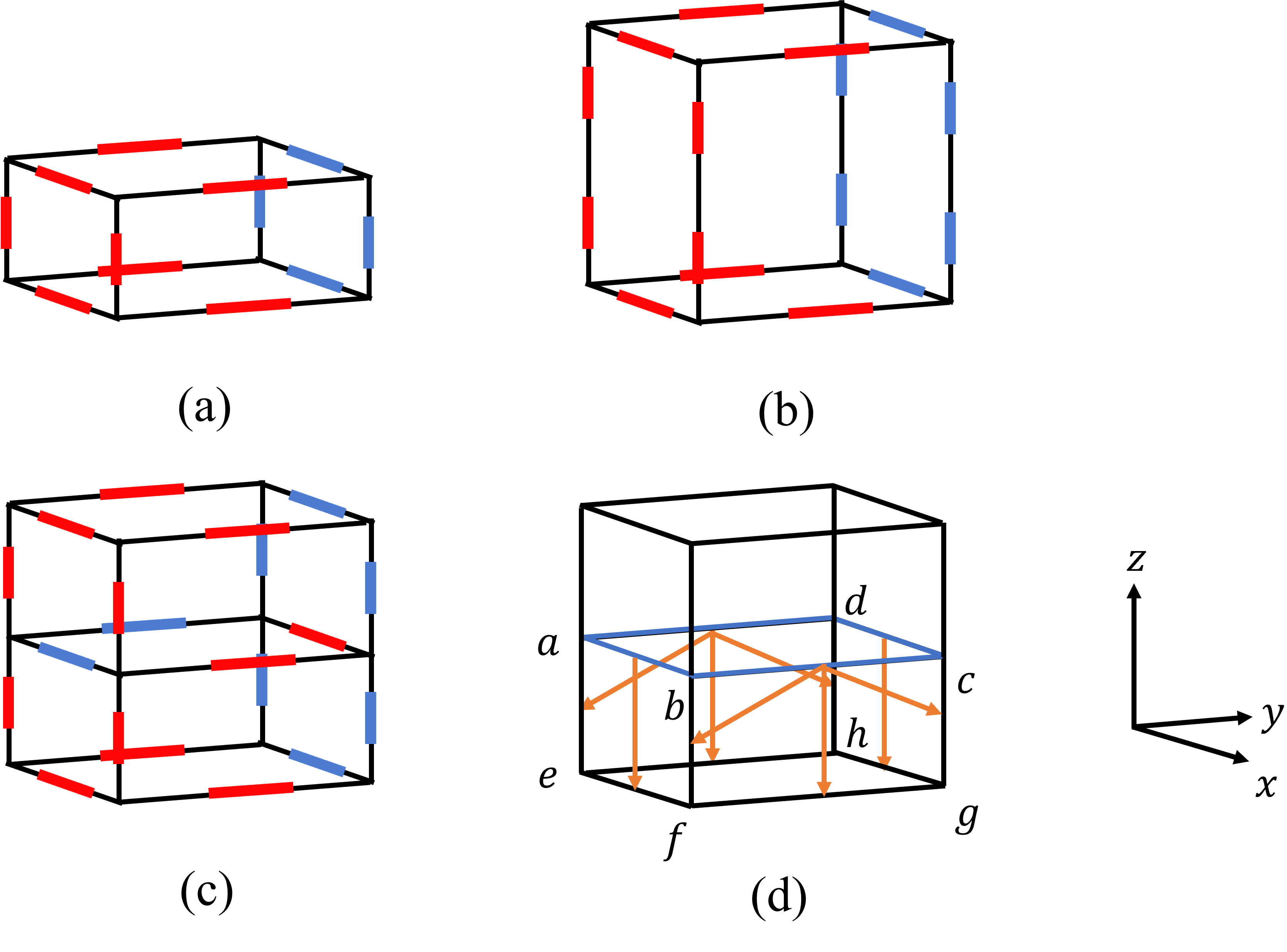}
	\caption{\textbf{ ERG transformation of the 3D X-cube model labeled by $[0,1,2,3]$.} This ERG transformation is denoted as $\mathtt{ERG}^1$ in Eq.~(\ref{eq_ERG_relation_Xcube}). We use red bars to denote spins occupied by strings in a configuration, and blue bars for the unoccupied ones. In (a), we demonstrate a configuration around the cube $\gamma_3=(\frac{1}{2},\frac{1}{2},\frac{1}{2})$ in the original $|\xi_i\ket$ state. In (b), we demonstrate the configuration obtained by cutting links of the form $(i,j,\frac{1}{2})$, rescaling the cut links, adding additional spins and applying a series of $\mathtt{CNOT}$ gates (i.e., a configuration from $|\xi_1\ket$). In (c), we demonstrate a configuration after further inserting a $[0,1,2,2]$ (2D toric code) state $|\xi_{gs}\ket$ (i.e., a configuration from $|\xi_2\ket=|\xi_1\ket \Otimes |\xi_{gs}\ket$). In (d), we demonstrate the $\mathtt{CNOT}$ gates applied on $|\xi_2\ket$. Here, we use a different notation for clarity. The control qubits, i.e., spins from the inserted $[0,1,2,2]$ state on plane $z=1$ (in the rescaled lattice), are denoted by blue links, while other spins are denoted by black links. The orange arrows point from control qubits to corresponding target qubits. Some vertices are denoted by letters.}
	\label{fig:xcubegates}
\end{figure}

In this subsection, we review the $\mathtt{ERG}^1$ transformation of the $[0,1,2,3]$ states following the recipe in Ref.~\cite{Shirley2018}. Again, we firstly give an intuitive picture of the $[0,1,2,3]$ states based on the general discussion in Sec.~\ref{sec:pre}. In $[0,1,2,3]$ (a.k.a. 3D X-cube) model, spins are located at links of a 3D cubic lattice. In this case, $3$ $B^l_j$ terms with perpendicular $l$, where $l$ denotes a plane containing vertex $j$, require $j$ can only emanate $3$ perpendicular strings composed of flipped spins (see Fig.~\ref{fig:xcubegates} (a)), thus flipped spins must form ``cages''~\cite{Prem2019a} in a superpositioned configuration. An $A_i$ term flips the $12$ spins on the links of cube $i$, thus contractible cages can freely fluctuate. We will see that, the ERG transformation indeed preserves this cage-net pattern of $[0,1,2,3]$ states. 

We start with a ground state $|\xi_i \rangle$ of  the $[0,1,2,3]$ model defined on a cubic lattice of the size $L_x\times L_y \times L_z$ with PBC, and obtain a ground state $|\xi_f \rangle$ on a cubic lattice of the size $L_x\times L_y \times (L_z + 1)$ with PBC by the following transformations:

First, we choose a $T^2$ ($2$-torus) composed of the centers of parallel links along direction $\hat{z}$ with the same $\hat{z}$-coordinate, and regard the $T^2$ as a cut: all links intersecting with the $T^2$ are cut into $2$ links. Without loss of generality, we assume the $T^2$ is located at $z=\frac{1}{2}$, which means the cut links are of the form $(i,j,\frac{1}{2})$, where $i,j$ are integers. After that, we apply a rescaling. For each cut link $l=(i,j,\frac{1}{2})$, we double the length of $l$ to $2$. Then, we can see that $l$ is cut into links $l_1=(i,j,\frac{1}{2})$ and $l_2=(i,j,\frac{3}{2})$ of length $1$, and now the cut $T^2$ is located at $z=1$. We assign the original spin on $l$ to $l_1$.
	
Second, for each $l_2$, we put an additional spin of the state $|0\ket$ on it. It means that we enlarge the Hilbert space by taking the tensor product of the original one and the added spins, and add a series of $-\sigma_{l_2}^z$ terms to the Hamiltonian to make all the added spins in the state $|0\ket$. Then, for each original cut link $l$, we apply a $\mathtt{CNOT}$ gate with the original qubit on $l_1$ as control qubit and the added one on $l_2$ as target. By conjugate action of $\mathtt{CNOT}$ gates, the added $-\sigma_{l_2}^z$ terms are mapped to $-\sigma_{l_1}^z \sigma_{l_2}^z$ terms (see Appendix~\ref{app_cnot}). As a result, given a cut link $l$, for an arbitrary Ising configuration $|\cdots \sigma_l \cdots \rangle$ from $|\xi_i \rangle$ ($\sigma_l = 0$ or $1$), we have $|\cdots \sigma_l \cdots \rangle \rightarrow |\cdots \sigma_{l_1} \sigma_{l_2} \cdots \rangle  $, where $\sigma_{l_1} =  \sigma_{l_2} = \sigma_{l}$ (see Fig.~\ref{fig:xcubegates} (b)). The ground state transformed by steps above is denoted as $|\xi_1\ket$.
	
Third, we insert a $[0,1,2,2]$ (2D toric code) state $|\xi_{gs}\rangle$ of the size $L_x\times L_y$ on the cut $T^2$ given in the first step. That is to say, the spins composing the inserted state are located on links of the form $(i+\frac{1}{2}, j, 1)$ and $(i, j+\frac{1}{2}, 1)$ in the rescaled lattice (see Fig.~\ref{fig:xcubegates} (c), note that there are no spins on such links before this step). Then, we denote the tensor product of $|\xi_1\rangle$ and $|\xi_{gs}\rangle$ as $|\xi_2\rangle = |\xi_1\rangle \Otimes |\xi_{gs}\rangle$. As $[0,1,2,2]$ (2D toric code) model on a $T^2$ is $4$-fold degenerated, this step has $4$ possible outcomes corresponding to $4$ possible inserted $[0,1,2,2]$ states.
	
Finally, we act a series of $\mathtt{CNOT}$ gates on $|\xi_2\rangle$ as illustrated in Fig.~\ref{fig:xcubegates} (d). The $\mathtt{CNOT}$ gates are organized in a translational invariant manner, thus we only need to specify them for a specific cube. Without loss of generality, we take $\gamma_3=(\frac{1}{2},\frac{1}{2},\frac{1}{2})$, and denote the vertices of $\gamma_3$ by letters as shown in Fig.~\ref{fig:xcubegates} (d). For example, we have $e=(0,0,0)$ and $c=(1,1,1)$. Then, the $\mathtt{CNOT}$ gates can be explicitly specified as follows:
\begin{align*}
	&\sigma_{bc} \rightarrow \sigma_{bf}, \sigma_{cg}, \sigma_{fg};\\
	&\sigma_{ad} \rightarrow \sigma_{ae}, \sigma_{dh}, \sigma_{eh};\\
	&\sigma_{ab} \rightarrow \sigma_{ef};\\
	&\sigma_{dc} \rightarrow \sigma_{hg};
\end{align*}
where $\sigma_{xy}$ refers to the spin located on the link between $x$ and $y$ vertices, $\rightarrow$ points from the control qubit to target qubits. Intuitively, we can see that by conjugate action (see Apppendix~\ref{app_cnot}), the $\mathtt{CNOT}$ gates map the $A_p=\sigma^x_{ab} \sigma^x_{bc} \sigma^x_{cd} \sigma^x_{da}$ stabilizer of the inserted $[0,1,2,2]$ state to $A_c=\sigma^x_{ab} \sigma^x_{bc} \sigma^x_{cd} \sigma^x_{da} \sigma^x_{ef} \sigma^x_{fg} \sigma^x_{gh} \sigma^x_{he}\sigma^x_{ae} \sigma^x_{bf} \sigma^x_{cg} \sigma^x_{dh}$, that is a stabilizer of $[0,1,2,3]$ (3D X-cube) state. Similarly, we can check that the $\mathtt{CNOT}$ gates generate all stabilizer generators we need to obtain $[0,1,2,3]$ states on a lattice of the size $L_x\times L_y \times (L_z + 1)$ with PBC (see Sec.~\ref{subsec:proof_ham} for a more detailed demonstration).
	
Therefore, after the application of the $\mathtt{CNOT}$ gates on $|\xi_2\rangle$, we obtain $|\xi_f\rangle$, which is a ground state of $[0,1,2,3]$ (3D X-cube) model on a lattice of the size $L_x\times L_y \times (L_z + 1)$ with PBC. Pictorially, we can see the transformed state preserves the cage-net pattern of $[0,1,2,3]$ states.
 
Due to the fact that there are $4$ possible choices of $|\xi_{gs}\rangle$ in the third step, for a given $|\xi_i \rangle$, we have $4$ possible $|\xi_f \rangle$ outcomes. As a result, if we require the GSD formula to be symmetric for $L_x$, $L_y$ and $L_z$, the GSD of $[0,1,2,3]$ model has to satisfy $\log_2 GSD = 2L_x + 2L_y + 2L_z + C$, where $C$ is a constant. This result is consistent with the exact result given in Ref.~\cite{Vijay2016,Li2021}. Besides, based on this method to obtain the GSD, it has been shown in Ref.~\cite{Shirley2018} that the coefficients of linear terms in the $\log_2 GSD$ are directly related to the topology of the 2D subsystems (dubbed as ``leaves'') of $[0,1,2,3]$ model. 

\subsection{Level-$2$ ERG transformation of $[0,1,2,4]$ states}
\label{subsec:0124_ER}

In this subsection, we demonstrate the $\mathtt{ERG}^2$ transformation of $[0,1,2,4]$ states. Similar to $[0,1,2,3]$ model, here we give an intuitive picture of $[0,1,2,4]$ states based on the general discussion in Sec.~\ref{sec:pre}. In $[0,1,2,4]$ model, spins are located at links of a 4D hypercubic lattice. In this case, $6$ $B^l_j$ terms with perpendicular $l$, where $l$ denotes a plane containing vertex $j$, require $j$ can only emanate $4$ perpendicular strings composed of flipped spins. An $A_i$ term flips the $32$ spins on the links of hypercube $i$. We will see that, the ERG transformation indeed preserves the pattern of $[0,1,2,4]$ states. 

Again, we start with a ground state $|\xi_i \rangle$ of the $[0,1,2,4]$ model defined on a lattice of the size $L_1\times L_2 \times L_3 \times L_4$ with PBC, and obtain a ground state $|\xi_f \rangle$ on a lattice of the size $L_1\times L_2 \times L_3 \times (L_4 + 1)$ with PBC. The $\mathtt{ERG}^2$ transformation can be described as follows:

First, we choose a $T^3$ ($3$-torus) composed of the centers of parallel links along direction $\hat{x}_4$ with the same $\hat{x}_4$-coordinate, and regard the $T^3$ as a cut: all links intersecting with the $T^3$ are cut into $2$ links. Without loss of generality, we assume the $T^3$ is located at $x_4=\frac{1}{2}$, which means the cut links are of the form $(i,j,k,\frac{1}{2})$, where $i,j,k$ are integers. After that, we apply a rescaling. For each cut link $l=(i,j,k,\frac{1}{2})$, we double the length of $l$ to $2$. Then, we can see that $l$ is cut into links $l_1=(i,j,k,\frac{1}{2})$ and $l_2=(i,j,k,\frac{3}{2})$ of length $1$, and now the cut $T^3$ is located at $x_4=1$. We assign the original spin on $l$ to $l_1$.
	
Second, for each $l_2$, we put an additional spin of the state $|0\ket$ on it. It means that we enlarge the Hilbert space by taking the tensor product of the original one and the added spins, and add a series of $-\sigma_{l_2}^z$ terms to the Hamiltonian to make all the added spins in the state $|0\ket$. Then, for each original cut link $l$, we apply a $\mathtt{CNOT}$ gate with the original qubit on $l_1$ as control qubit and the added one on $l_2$ as target. By conjugate action of $\mathtt{CNOT}$ gates, the added $-\sigma_{l_2}^z$ terms are mapped to $-\sigma_{l_1}^z \sigma_{l_2}^z$ terms (see Appendix~\ref{app_cnot}). As a result, given a cut link $l$, for an arbitrary Ising configuration $|\cdots \sigma_l \cdots \rangle$ from $|\xi_i \rangle$ ($\sigma_l = 0$ or $1$), we have $|\cdots \sigma_l \cdots \rangle \rightarrow |\cdots \sigma_{l_1} \sigma_{l_2} \cdots \rangle  $, where $\sigma_{l_1} =  \sigma_{l_2} = \sigma_{l}$. The ground state transformed by the steps above is denoted as $|\xi_1\ket$.
	
Third, we insert a $[0,1,2,3]$ (3D X-cube) state $|\xi_{gs}\rangle$ of the size $L_1\times L_2 \times L_3$ on the cut $T^3$ given in the first step. That is to say, the spins composing the inserted state are located on links of the form $(i+\frac{1}{2}, j, k, 1)$, $(i, j+\frac{1}{2}, k, 1)$ and $(i,j,k+\frac{1}{2},1)$ in the rescaled lattice (note that there are no spins on such links before this step). Then, we denote the tensor product of $|\xi_1\rangle$ and $|\xi_{gs}\rangle$ as $|\xi_2\rangle = |\xi_1\rangle \Otimes |\xi_{gs}\rangle$. As $[0,1,2,3]$ (3D X-cube) model on the $T^3$ satisfies $\log_2 GSD= 2L_1+ 2L_2+ 2L_3 - 3$, this step has $2^{2L_1+ 2L_2+ 2L_3 - 3}$ possible outcomes corresponding to $2^{2L_1+ 2L_2+ 2L_3 - 3}$ possible inserted $[0,1,2,3]$ states.
	
\begin{figure}
	\centering
	\includegraphics[width=1.0\linewidth]{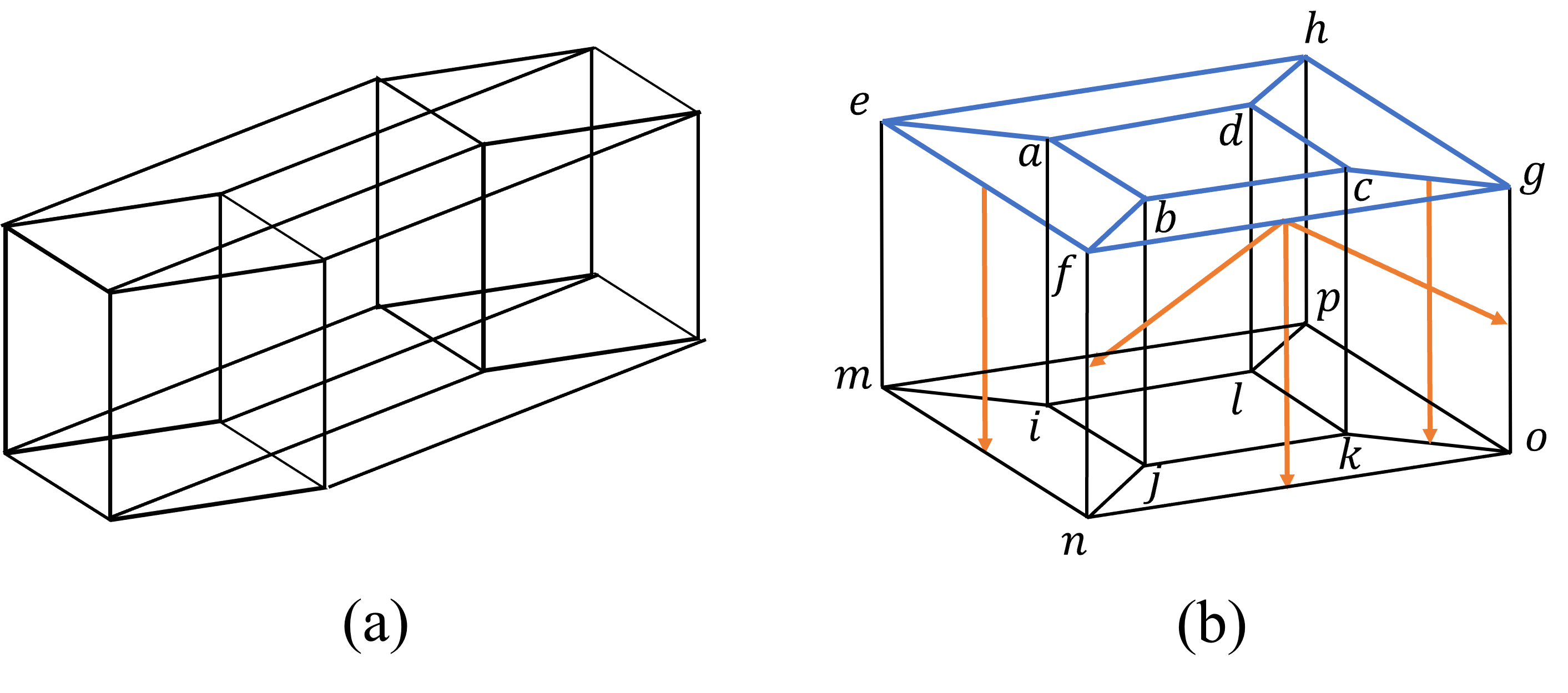}
	\caption{\textbf{ ERG transformation of the $[0,1,2,4]$ model.} Here we demonstrate the $\CNOT$ gates applied on $|\xi_2\ket$ following the same rules as in Fig.~\ref{fig:xcubegates} (d). In (a), we give a schematic picture of a $4$-cube, while in (b), we use another way to illustrate the $4$-cube to show the $\CNOT$ gates more clearly. In (b), we use blue links to denote control qubits from the inserted $[0,1,2,3]$ (3D X-cube) state on the cut $T^3$ with $x_4=1$, and black links for the target qubits. For simplicity, orange arrows pointing from control qubits to target qubits are only presented for $3$ control qubits along different directions. As we can see, these $\CNOT$ gates satisfy all conditions given in the general recipe in Sec.~\ref{subsec:proof_ham}. }
	\label{fig:0124gates}
\end{figure}

Finally, we act a series of $\mathtt{CNOT}$ gates on $|\xi_2\rangle$ as illustrated in Fig.~\ref{fig:0124gates}. The $\mathtt{CNOT}$ gates are organized in a translational invariant manner, thus we only need to specify them for a specific $4$-cube. Without loss of generality, we take $\gamma_4=(\frac{1}{2},\frac{1}{2},\frac{1}{2},\frac{1}{2})$, and denote the vertices of $\gamma_4$ by letters as shown in Fig.~\ref{fig:0124gates}. Then, the $\mathtt{CNOT}$ gates can be explicitly specified as follows:
	\begin{align*}
		&\sigma_{fg} \rightarrow \sigma_{fn}, \sigma_{no}, \sigma_{og};\\
		&\sigma_{bc} \rightarrow \sigma_{bj}, \sigma_{jk}, \sigma_{kc};\\
		&\sigma_{ad} \rightarrow \sigma_{ai}, \sigma_{il}, \sigma_{ld};\\
		&\sigma_{eh} \rightarrow \sigma_{em}, \sigma_{mp}, \sigma_{ph};\\
		&\sigma_{ef} \rightarrow \sigma_{mn};\\
		&\sigma_{ab} \rightarrow \sigma_{ij};\\
		&\sigma_{dc} \rightarrow \sigma_{lk};\\
		&\sigma_{hg} \rightarrow \sigma_{po};\\
		&\sigma_{cg} \rightarrow \sigma_{ko};\\
		&\sigma_{bf} \rightarrow \sigma_{jn};\\
		&\sigma_{ae} \rightarrow \sigma_{im};\\
		&\sigma_{dh} \rightarrow \sigma_{lp};\\
	\end{align*}
	where $\sigma_{xy}$ refers to the spin located on the link between $x$ and $y$ vertices, $\rightarrow$ points from the control qubit to target qubits. Intuitively, we can see that by conjugate action (see Apppendix~\ref{app_cnot}), the $\mathtt{CNOT}$ gates map the 
	\begin{align*}
		A_c=\sigma^x_{ab} \sigma^x_{bc} \sigma^x_{cd} \sigma^x_{da} \sigma^x_{ae} \sigma^x_{bf} \sigma^x_{cg} \sigma^x_{dh} \sigma^x_{ef} \sigma^x_{fg} \sigma^x_{gh} \sigma^x_{he},
	\end{align*}
 	stabilizer of the inserted $[0,1,2,3]$ state to
	\begin{align*}
		A_{\gamma_4}=&\sigma^x_{ab} \sigma^x_{bc} \sigma^x_{cd} \sigma^x_{da} \sigma^x_{ae} \sigma^x_{bf} \sigma^x_{cg} \sigma^x_{dh}\\
		&\sigma^x_{ef} \sigma^x_{fg} \sigma^x_{gh} \sigma^x_{he} \sigma^x_{ij} \sigma^x_{jk} \sigma^x_{kl} \sigma^x_{li}\\ 
		&\sigma^x_{im} \sigma^x_{jn} \sigma^x_{ko} \sigma^x_{lp} \sigma^x_{mn} \sigma^x_{no} \sigma^x_{op} \sigma^x_{pm}\\ 
		&\sigma^x_{ai} \sigma^x_{bj} \sigma^x_{ck} \sigma^x_{dl} \sigma^x_{em} \sigma^x_{fn} \sigma^x_{go} \sigma^x_{hp},
	\end{align*} that is a stabilizer of $[0,1,2,4]$ ground state. Similarly, we can check that the $\mathtt{CNOT}$ gates generate all stabilizer generators we need to obtain $[0,1,2,4]$ ground states on a lattice of the size $L_1\times L_2 \times L_3 \times (L_4 + 1)$ with PBC (see Sec.~\ref{subsec:proof_ham} for a more detailed demonstration). Therefore, after the application of the $\mathtt{CNOT}$ gates on $|\xi_2\rangle$, we obtain $|\xi_f\rangle$, which is a ground state of $[0,1,2,4]$ model on a lattice of the size $L_1\times L_2 \times L_3 \times (L_4 + 1)$ with PBC.

Similar to $[0,1,2,3]$ model, we can see that the GSD of $[0,1,2,4]$ model has to satisfy $\log_2 GSD = (2L_1+2L_2+2L_3-3)L_4 + C(L_1, L_2, L_3)$, where $C(L_1, L_2, L_3)$ is a function of $L_1$, $L_2$ and $L_3$. When we require the GSD formula to be symmetric for $L_1$, $L_2$, $L_3$ and $L_4$, then we have $\log_2 GSD = 2L_1 L_2 + 2L_1 L_3 +2 L_1 L_4 + 2L_2 L_3 + 2L_2 L_4 + 2L_3 L_4 - 3L_1 - 3L_2 - 3L_3- 3L_4 + C'$, where $C'$ is a constant. This result is consistent with the result obtained by ground state decomposition in Ref.~\cite{Li2021}.

\subsection{Level-$1$ ERG transformation of $[1,2,3,4]$ states}
\label{subsec:1234_ER}

For comparison, in this subsection, we demonstrate the $\mathtt{ERG}^1$ transformation of $[1,2,3,4]$ states. We will see that, though $[1,2,3,4]$ model has the same spatial dimension as $[0,1,2,4]$ model, in the ERG transformation of its ground states we only need to add/remove $\mathtt{LRE}^1$ rather than $\mathtt{LRE}^2$ states. Before the demonstration, here we also give an intuitive picture of $[1,2,3,4]$ states based on the general discussion in Sec.~\ref{sec:pre}. In $[1,2,3,4]$ model, spins are located at plaquettes of a 4D hypercubic lattice. In this case, $3$ $B^l_j$ terms with perpendicular $l$, where $l$ denotes a 3D subsystem containing link $j$, require $j$ can only emanate $3$ perpendicular membranes composed of flipped spins. An $A_i$ term flips the $24$ spins on the plaquettes of hypercube $i$. We will see that, the ERG transformation indeed preserves the pattern of $[1,2,3,4]$ states. 

Again, we start with a ground state $|\xi_i \rangle$ of the $[1,2,3,4]$ model defined on a lattice of the size $L_1\times L_2 \times L_3 \times L_4$ with PBC, and obtain such a ground state $|\xi_f \rangle$ on a lattice of the size $L_1\times L_2 \times L_3 \times (L_4 + 1)$ with PBC. The $\mathtt{ERG}^1$ transformation can be similarly described as follows:

First, we choose a $T^3$ ($3$-torus) composed of the centers of parallel links along direction $\hat{x}_4$ with the same $\hat{x}_4$-coordinate, and regard the $T^3$ as a cut: all plaquettes intersecting with the $T^3$ are cut into $2$ plaquettes. Without loss of generality, we assume the $T^3$ is located at $x_4=\frac{1}{2}$, which means the cut plaquettes are of the form $(i,j,k,\frac{1}{2})+\frac{1}{2} I_n$, where $n=1,2,3$, $I_n$ is the unit vector along $\hat{x}_n$ direction, $i,j,k$ are integers. After that, we apply a rescaling. For each cut plaquette $p=(i,j,k,\frac{1}{2})+\frac{1}{2} I_n$, we double the linear size of $p$ along $\hat{x}_4$ direction to $2$. Then, we can see that $p$ is cut into plaquettes $p_1=(i,j,k,\frac{1}{2})+\frac{1}{2} I_n$ and $p_2=(i,j,k,\frac{3}{2})+\frac{1}{2} I_n$ with linear sizes along $\hat{x}_4$ direction equal to $1$, and now the cut $T^3$ is located at $x_4=1$. We can assign the original spin on $p$ to $p_1$.
	
Second, for each $p_2$, we put an additional spin of the state $|0\ket$ on it. Equivalently, it means that we enlarge the Hilbert space by taking the tensor product of the original one and the added spins, and add a series of $-\sigma_{p_2}^z$ terms to the Hamiltonian to make all the added spins in the state $|0\ket$. Then, for each original cut plaquette $p$, we apply a $\mathtt{CNOT}$ gate with the original qubit on $p_1$ as control qubit and the added one on $p_2$ as target. By conjugate action of $\mathtt{CNOT}$ gates, the added $-\sigma_{p_2}^z$ terms are mapped to $-\sigma_{p_1}^z \sigma_{p_2}^z$ terms (see Appendix~\ref{app_cnot}). As a result, given a cut plaquette $p$, for an arbitrary Ising configuration $|\cdots \sigma_p \cdots \rangle$ from $|\xi_i \rangle$ ($\sigma_p = 0$ or $1$), we have $|\cdots \sigma_p \cdots \rangle \rightarrow |\cdots \sigma_{p_1} \sigma_{p_2} \cdots \rangle$, where $\sigma_{p_1} =  \sigma_{p_2} = \sigma_{p}$. The ground state transformed by the steps above is denoted as $|\xi_1\ket$.
	
Third, we insert a $[1,2,3,3]$ (3D toric code) state $|\xi_{gs}\rangle$ of the size $L_1\times L_2 \times L_3$ on the cut $T^3$ given in the first step. That is to say, the spins composing the inserted state are located on plaquettes of the form $(i+\frac{1}{2}, j+\frac{1}{2}, k, 1)$, $(i, j+\frac{1}{2}, k+\frac{1}{2}, 1)$ and $(i+\frac{1}{2},j,k+\frac{1}{2},1)$ in the rescaled lattice (note that there are no spins on such plaquettes before this step). Then, we denote the tensor product of $|\xi_1\rangle$ and $|\xi_{gs}\rangle$ as $|\xi_2\rangle = |\xi_1\rangle \Otimes |\xi_{gs}\rangle$. As $[1,2,3,3]$ (3D toric code) model on the $T^3$ satisfies $\log_2 GSD= 3~$\cite{Hamma2005,Kong2020}, this step has $2^{3}$ possible outcomes corresponding to $2^{3}$ possible inserted $[1,2,3,3]$ states.
	
\begin{figure*}
	\centering
	\includegraphics[width=0.9\linewidth]{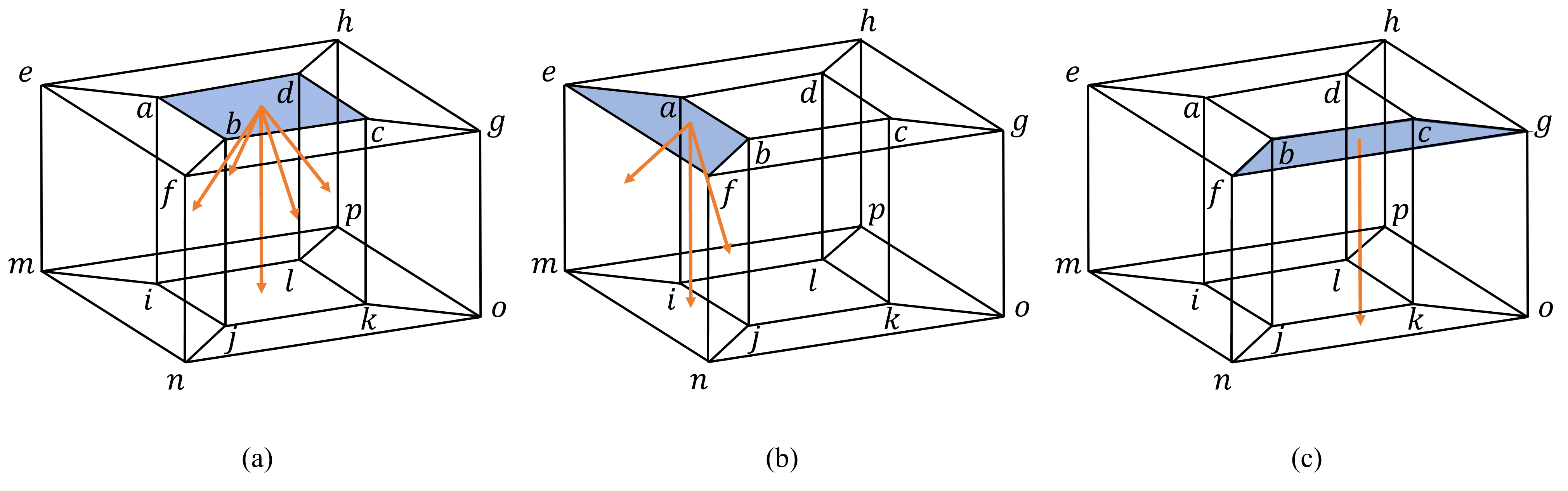}
	\caption{\textbf{ERG transformation of the $[1,2,3,4]$ model.} Here we follow a similar notation as in Fig.~\ref{fig:0124gates} (b) to demonstrate the $\mathtt{CNOT}$ gates applied on $|\xi_2\ket$. Again, we set qubits nearest to the cube specified by $abcdefgh$ as from the inserted $[1,2,3,3]$ (3D toric code) state on the cut $T^3$ with $x_4=1$. For clarity, here we only demonstrate $3$ control qubits and their associated $\mathtt{CNOT}$ gates. In (a), (b) and (c), we use three transparent plaquettes highlighted with blue to denote three different control qubits, and present orange arrows pointing from control qubits to target ones. Here it should be noticed that two different plaquettes may share the same center in these pictures, like $abcd$ and $efgh$. As we can see, these $\CNOT$ gates satisfy all conditions given in the general recipe in Sec.~\ref{subsec:proof_ham}.}
	\label{fig:1234gates}
\end{figure*}

Finally, we act a series of $\mathtt{CNOT}$ gates on $|\xi_2\rangle$ as illustrated in Fig.~\ref{fig:1234gates}. The $\mathtt{CNOT}$ gates are organized in a translational invariant manner, thus we only need to specify them for a specific $4$-cube. Without loss of generality, we take $\gamma_4=(\frac{1}{2},\frac{1}{2},\frac{1}{2},\frac{1}{2})$, and denote the vertices of $\gamma_4$ by letters as shown in Fig.~\ref{fig:1234gates}. Then, the $\mathtt{CNOT}$ gates can be explicitly specified as follows:
	\begin{align*}
		&\sigma_{abcd} \rightarrow \sigma_{ijkl}, \sigma_{abji}, \sigma_{bckj}, \sigma_{cdlk}, \sigma_{dail};\\
		&\sigma_{efgh} \rightarrow \sigma_{mnop}, \sigma_{efnm}, \sigma_{fgon}, \sigma_{ghpo}, \sigma_{hemp};\\
		&\sigma_{abfe} \rightarrow \sigma_{ijnm}, \sigma_{aemi}, \sigma_{bfnj};\\
		&\sigma_{cdhg} \rightarrow \sigma_{klpo}, \sigma_{dhpl}, \sigma_{cgok};\\
		&\sigma_{bcgf} \rightarrow \sigma_{jkon};\\
		&\sigma_{daeh} \rightarrow \sigma_{limp};\\
	\end{align*}
	where $\sigma_{xyzw}$ refers to the spin located on the plaquette between $x$, $y$, $z$ and $w$ vertices, $\rightarrow$ points from the control qubit to target qubits. Intuitively, we can see that by conjugate action (see Apppendix~\ref{app_cnot}), the $\mathtt{CNOT}$ gates map the 
	\begin{align*}
		A_c=\sigma^x_{abcd} \sigma^x_{efgh} \sigma^x_{abfe} \sigma^x_{bcgf} \sigma^x_{cdhg} \sigma^x_{daeh},
	\end{align*}
	stabilizer of the inserted $[1,2,3,3]$ state to
	\begin{align*}
		A_{\gamma_4}=&\sigma^x_{abcd} \sigma^x_{efgh} \sigma^x_{abfe} \sigma^x_{bcgf} \sigma^x_{cdhg} \sigma^x_{daeh}\\
		&\sigma^x_{ijkl} \sigma^x_{mnop} \sigma^x_{ijnm} \sigma^x_{jkon} \sigma^x_{klpo} \sigma^x_{limp}\\ 
		&\sigma^x_{aemi} \sigma^x_{bfnj} \sigma^x_{efnm} \sigma^x_{abji} \sigma^x_{bckj} \sigma^x_{fgon} \\ 
		&\sigma^x_{cgok} \sigma^x_{dhpl} \sigma^x_{ghpo} \sigma^x_{cdlk} \sigma^x_{dail} \sigma^x_{hemp} ,
	\end{align*} that is a stabilizer of $[1,2,3,4]$ ground state. Similarly, we can check that the $\mathtt{CNOT}$ gates generate all stabilizer generators we need to obtain $[1,2,3,4]$ ground states on a lattice of the size $L_1\times L_2 \times L_3 \times (L_4 + 1)$ with PBC (see Sec.~\ref{subsec:proof_ham} for a more detailed demonstration). Therefore, after the application of the $\mathtt{CNOT}$ gates on $|\xi_2\rangle$, we obtain $|\xi_f\rangle$, which is a ground state of $[1,2,3,4]$ model on a lattice of the size $L_1\times L_2 \times L_3 \times (L_4 + 1)$ with PBC.

Similar to $[0,1,2,3]$ model, we can see that the GSD of $[1,2,3,4]$ model has to satisfy $\log_2 GSD = 3\times L_4 + C(L_1,L_2,L_3)$, where $C(L_1,L_2,L_3)$ is a function of $L_1,L_2$ and $L_3$. When we require the GSD formula to be symmetric for $L_1$, $L_2$, $L_3$ and $L_4$, then we have $\log_2 GSD = 3L_1 + 3L_2 + 3L_3 + 3L_4 + C'$, where $C'$ is a constant. This result is consistent with the result obtained by ground state decomposition in Ref.~\cite{Li2021}.

\section{ERG of generic levels}
\label{sec:general}

In this section, we firstly show a generic recipe of the construction of $\mathtt{ERG}^{D-d-2}$ transformations of $[d,d+1,d+2,D]$ models with $D > d+2$.  After that, in Sec.~\ref{subsec:proof_ham}, we prove that for such a $[d,d+1,d+2,D]$ model, the constructed ERG transformation indeed gives ground states of the same model of different sizes, i.e., the models are fixed points of the corresponding ERG transformations. Finally, in Sec.~\ref{subsec:dis}, we discuss about the hierarchy of ERG transformations and LRE states based on the constructed ERG transformations. Note that $\ERG^0$ transformations of $[D-2,D-1,D,D]$ models are not included in this recipe.

\subsection{Level-$(D-d-2)$ ERG transformation of $[d,d+1,d+2,D]$ states}
\label{subsec:dddD_ER}

In general, for a $[d,d+1,d+2,D]$ model with $D > d+2$, we can demonstrate the $\mathtt{ERG}^{D-d-2}$ transformation of $[d,d+1,d+2,D]$ states. Again, we start with a ground state $|\xi_i \rangle$ of $[d,d+1,d+2,D]$ model defined on a lattice of the size $L_1\times L_2 \times \cdots \times L_D$ with PBC, and obtain a ground state $|\xi_f \rangle$ on a lattice of the size $L_1\times L_2 \times \cdots \times (L_D + 1)$ with PBC. The $\mathtt{ERG}^{D-d-2}$ transformation can be described as follows:

\begin{enumerate}
	
	\item[$-$] First, we choose a $(D-1)$-torus $T^{D-1}$ composed of the centers of links with the same $\hat{x}_D$-coordinate. Without loss of generality, we set the chosen $T^{D-1}$ to be located at $x_D=\frac{1}{2}$, such that it is composed of the centers of links of the form $(n_1,n_2,\cdots,n_{D-1},\frac{1}{2})$, where $n_1,n_2,\cdots,n_{D-1}$ are integers. Then we regard the $T^{D-1}$ as a cut: every $\gamma_{d+1}$ intersecting with the $T^{D-1}$ is cut into $2$ $\gamma_{d+1}$'s with identical spins. That is to say, for each cut $\gamma_{d+1}$, we put an additional spin in the state $|0\ket$, and then apply a $\mathtt{CNOT}$ gate with the original qubit as control qubit and the added one as target. In consequence, given a cut $\gamma_{d+1}$, for an arbitrary Ising configuration $|\cdots \sigma_{\gamma_{d+1}} \cdots \rangle$ from $|\xi_i \rangle$ ($\sigma_{\gamma_{d+1}} = 0$ or $1$), we have $|\cdots \sigma_{\gamma_{d+1}} \cdots\ket \rightarrow |\cdots \sigma_{({\gamma_{d+1}})_1}\sigma_{({\gamma_{d+1}})_2} \cdots \ket$, where $\sigma_{({\gamma_{d+1}})_1} =  \sigma_{({\gamma_{d+1}})_2} = \sigma_{\gamma_{d+1}}$. Then, we rescale the lattice by extending the linear size of the cut $\gamma_{d+1}$ along $\hat{x}_D$ direction to $2$, such that now the chosen $T^{D-1}$ is composed of sites of the form $(n_1,n_2,\cdots,n_{D-1},1)$, and for a cut $\gamma_{d+1}=(\cdots,\frac{1}{2})$ in the original lattice, the original and additional spins are respectively assigned to $({\gamma_{d+1}})_1 = (\cdots,\frac{1}{2})$ and $({\gamma_{d+1}})_2 = (\cdots,\frac{3}{2})$ in the rescaled lattice. The ground state transformed by this step is denoted as $|\xi_1\rangle$.
	
	\item[$-$] Second, we put a $[d,d+1,d+2,D-1]$ ground state $|\xi_{gs}\rangle$ of the size $L_1\times \cdots \times L_{D-1}$ on the $T^{D-1}$ given in the previous step. That is to say, we can regard the $(n_1,n_2,\cdots,n_{D-1}, 1)$ sites as forming a hypercubic lattice defined on the $T^{D-1}$, and consider a $[d,d+1,d+2,D-1]$ ground state $|\xi_{gs}\rangle$ defined on this lattice. Then, by taking the tensor product of $|\xi_1\rangle$ and $|\xi_{gs}\rangle$, we obtain $|\xi_2\rangle = |\xi_1\rangle \Otimes |\xi_{gs}\rangle$.
	
	\item[$-$] Third, we act an LU transformation $\mathcal{U}$ composed of a series of $\mathtt{CNOT}$ gates on $|\xi_2\rangle$ (see Sec.~\ref{subsec:proof_ham} for a demonstration of this LU transformation $\mathcal{U}$). After that, we obtain $|\xi_f\rangle$, which is a ground state of $[d,d+1,d+2,D]$ model on a lattice of the size $L_1\times L_2 \times \cdots \times (L_D + 1)$ with PBC. 
\end{enumerate}
	
To see that this generic recipe is consistent with the GSD results obtained by ground state decomposition in Ref.~\cite{Li2021}, without loss of generality, say that in the polynomial $\log_2 GSD$ of $[d,d+1,d+2,D-1]$ model on the $T^{D-1}$, the coefficient of $L_i L_j \cdots L_n$ term is $c$ (here $i<j<\cdots<n<D$ is assumed). Then, the above ERG transformation requires that the number of copies of $c L_i L_j \cdots L_n$ contained in the $\log_2 GSD$ of $[d,d+1,d+2,D]$ model grows linearly with $L_D$. That is to say, the polynomial $\log_2 GSD$ of the $[d,d+1,d+2,D]$ model has to contain the term $cL_i L_j \cdots L_n L_D$. This result is consistent with the relevant results from Ref.~\cite{Li2021}.

\subsection{$[d,d+1,d+2,D]$ models as fixed points of level-$(D-d-2)$ ERG transformations}
\label{subsec:proof_ham}

In this subsection, we give the conditions that an LU transformation $\mathcal{U}$ used in the Step 3 of the $\mathtt{ERG}^{D-d-2}$ transformation of a general $[d,d+1,d+2,D]$ state should satisfy, and prove that the such an LU transformation $\mathcal{U}$ indeed gives ground states of the $[d,d+1,d+2,D]$ model on a lattice of different sizes, by considering the conjugate action of $\mathcal{U}$ on the Hamiltonian terms. Without loss of generality, we assume the cut $T^{D-1}$ is extended along $\hat{x}_1$, $\hat{x}_2$, $\cdots, \hat{x}_{D-1}$ directions, and the location is given by $x_D = 1$ (in the rescaled lattice). For convenience, here we explicitly write down the Hamiltonian of a $[d,d+1,d+2,D]$ model as below:
\begin{align}
	\label{eq:dddD}
	H_{[d,d+1,d+2,D]} = -\sum_{\gamma_D} A_{\gamma_D} - \sum_{\gamma_d} \sum_l B^l_{\gamma_d}, 
\end{align}
where an $A_{\gamma_D}$ term is the product of the $x$-components of the $\binom{D}{d+1} \cdot 2^{D-d-1}$ spins nearest to the $\gamma_D$, a $B^l_{\gamma_d}$ term is the product of the $z$-components of the $4$ spins that are (a) nearest to the $\gamma_d$ and (b) living in the $(d+2)$-dimensional subsystem $l$.

We start with the conditions that the LU transformation $\mathcal{U}$ should satisfy. According to the LU transformation $\mathcal{U}$ of $[0,1,2,3]$, $[0,1,2,4]$ and $[1,2,3,4]$ models (i.e., the $\CNOT$ gates applied on $|\xi_2\ket$ states of corresponding subsections), we expect such an LU transformation in the $\mathtt{ERG}^{D-d-2}$ transformation of a general $[d,d+1,d+2,D]$ state to satisfy the following conditions:
\begin{itemize}
	\item[$-$] First, we require $\mathcal{U}$ to be composed of a series of $\mathtt{CNOT}$ gates around $\gamma_D$'s with $x_D=\frac{1}{2}$, where all control qubits are from the cut $T^{D-1}$ (i.e., being located on $\gamma_{d+1}$'s with $x_D=1$). Besides, we require $\mathcal{U}$ to be translational invariant, such that the application of the $\mathtt{CNOT}$ gates is the same for every applied $\gamma^D$. Therefore, we only need to consider the application of $\mathtt{CNOT}$ gates in a single $\gamma_D$ to specify $\mathcal{U}$. Without loss of generality, we can focus on  $\gamma^r_D=(\frac{1}{2},\frac{1}{2},\cdots,\frac{1}{2})$. Here the superscript $r$ is for reference.
	\item[$-$] Second, for each $\gamma_{d+1}$ in $\gamma^r_D$ with $x_D=0$, we require the qubit on it to be controlled by the qubit on $\gamma_{d+1} + I_D$, where $I_D = (0,0,\cdots,0,1)$ is the unit vector along $\hat{x}_D$ direction. Obviously, the qubit on such a $\gamma_{d+1}$ is only controlled by $1$ control qubit in $\gamma^r_D$.
	\item[$-$] Third, for each $\gamma_{d+1}$ in $\gamma^r_D$ with $x_D=\frac{1}{2}$, we require the qubit on it to be controlled by exactly $1$ nearest control qubit in $\gamma^r_D$. Besides, for a pair of nearest parallel qubits, we require them to be simultaneously controlled (or not) by the control qubit that links them. For example, a pair of nearest parallel qubits respectively defined on $(\underbrace{\frac{1}{2},\frac{1}{2},\cdots,\frac{1}{2}}_{d},\underbrace{0,0,\cdots,0,\frac{1}{2}}_{D-d})$ and $(\underbrace{\frac{1}{2},\frac{1}{2},\cdots,\frac{1}{2}}_{d},\underbrace{1,0,\cdots,0,\frac{1}{2}}_{D-d})$ are either both controlled by $(\underbrace{\frac{1}{2},\frac{1}{2},\cdots,\frac{1}{2}}_{d},\underbrace{\frac{1}{2},0,\cdots,0,1}_{D-d})$ or not (here we can notice that this control qubit is the only one that links the pair, i.e., simultaneously being nearest to the pair of qubits).
\end{itemize}
The existence of such LU transformations is obvious. And we can check that when (a) $d=0,D=3$, (b) $d=0,D=4$ and (c) $d=1,D=4$, the LU transformations $\mathcal{U}$ in the ERG transformations of $[0,1,2,3]$, $[0,1,2,4]$ and $[1,2,3,4]$ states all satisfy the above conditions. Besides, here we should notice that each target qubit $\sigma_i$ with $x_D=\frac{1}{2}$ is always controlled by two qubits. Without loss of generality, say the qubit $\sigma_i$ on $(d+1)$-cube $i=(\underbrace{\frac{1}{2},\frac{1}{2},\cdots,\frac{1}{2}}_{d},\underbrace{0,0,\cdots,0,\frac{1}{2}}_{D-d})$ is controlled by the qubit on $i_c=(\underbrace{\frac{1}{2},\frac{1}{2},\cdots,\frac{1}{2}}_{d},\underbrace{\frac{1}{2},0,\cdots,0,1}_{D-d})$, according to the translational invariance of the LU transformation, the qubit on $i' = i - I_{d+1}=(\underbrace{\frac{1}{2},\frac{1}{2},\cdots,\frac{1}{2}}_{d},\underbrace{-1,0,\cdots,0,\frac{1}{2}}_{D-d})$ must be controlled by the qubit on $i_{c}'= i_c - I_{d+1} = (\underbrace{\frac{1}{2},\frac{1}{2},\cdots,\frac{1}{2}}_{d},\underbrace{-\frac{1}{2},0,\cdots,0,1}_{D-d})$; after that, as $i$ and $i'$ are parallel $(d+1)$-cubes connected by $i_{c}'$, $\sigma_i$ must also be controlled by the qubit on $i_{c}'$. Then, we can notice that an arbitrary $\gamma_D$ nearest to $i$ has the form $(\underbrace{\frac{1}{2},\frac{1}{2},\cdots,\frac{1}{2}}_{d},\underbrace{\pm \frac{1}{2},\pm \frac{1}{2},\cdots,\pm \frac{1}{2}, \frac{1}{2}}_{D-d})$, thus it must be either nearest to $i_c$ or $i_{c}'$. Since a target qubit can only be controlled by $1$ control qubit from a nearest $\gamma_D$ as required by the conditions above, no other qubits in the $T^{D-1}$ can control $\sigma_i$. In conclusion, for any target qubit $\sigma_i$ with $x_D=\frac{1}{2}$, there are always $2$ qubits that control it.

Then, we show that though the concrete form of the LU transformation $\mathcal{U}$ has not been specified, the above conditions can make sure that $\mathcal{U}$ produces the ground states as expected. That is to say, for an LU transformation $\mathcal{U}$ satisfying the conditions above, a $|\xi_f\ket=\mathcal{U}|\xi_2\ket$ is indeed a ground state of $[d,d+1,d+2,D]$ model. 

At first, we notice that $\mathcal{U}$ is applied on the $|\xi_2\rangle$ given in Sec.~\ref{subsec:dddD_ER}, and $|\xi_2\rangle$ can be obtained as a ground state of the following Hamiltonian:
\begin{align}
	H_1 = H_{dddD} + H_{dddD-1} + H_{zz},
\end{align}
where $H_{dddD}$ refers to the terms in the original $[d,d+1,d+2,D]$ model with some modifications according to the cut $\gamma_{d+1}$'s (see below), $H_{dddD-1}$ refers to the terms of the $[d,d+1,d+2,D-1]$ Hamiltonian on the cut $T^{D-1}$, and $H_{zz} = -\sum_i \sigma^z_i \sigma^z_{i+I_D}$ is added to make each pair of spins on a cut $\gamma_{d+1}$ identical, where $i$ refers to a $(d+1)$-cube with $x_D=\frac{1}{2}$ in the rescaled lattice. Note that the $A_{\gamma_D}$ terms in $H_{dddD}$ near the $T^{D-1}$ are modified to $A_{\gamma_D}'=A_{\gamma_D} A_{\gamma_D+I_D}$ to be consistent with the cut $\gamma_{d+1}$'s, where $A_{\gamma_D}$ and $A_{\gamma_D+I_D}$ have the same form as an ordinary $A$ term from the original $[d,d+1,d+2,D]$ model, and $\gamma_D$ satisfies $x_D=\frac{1}{2}$. As a concrete example, in $[0,1,2,4]$ model, where $d=0$, $D=4$, such a modified $A_{\gamma_4}$, denoted as $A_{\gamma_4}'$, is given by $A_{\frac{1}{2}, \frac{1}{2},\frac{1}{2}, \frac{1}{2}}' = A_{\frac{1}{2}, \frac{1}{2},\frac{1}{2}, \frac{1}{2}} A_{\frac{1}{2}, \frac{1}{2},\frac{1}{2}, \frac{3}{2}}$, where operators $A_{\frac{1}{2}, \frac{1}{2},\frac{1}{2}, \frac{1}{2}}$ and $A_{\frac{1}{2}, \frac{1}{2},\frac{1}{2}, \frac{3}{2}}$ themselves do not present in $H_{dddD}$. Furthermore, for a $B_{\gamma_d}$ term from $H_{dddD}$ with $x_D=2$ that involves qubit $\sigma_i$ with $x_D=\frac{1}{2}$, we can replace it by the product of the $B$ term itself and a corresponding $\sigma^z_i \sigma^z_{i+I_D}$ term, such that the $\sigma^z_i$ in the $B$ term is replaced by $\sigma^z_{i+I_D}$. This modification makes such $B$ terms ``connected''. For example, in $[0,1,2,3]$ model, due to our assignment that for a cut link the original qubit is put on a link of the form $(\cdots,\frac{1}{2})$, in the rescaled lattice, we would have $B$ terms such as $B^x_{(0,0,2)}=\sigma^z_{(0,\frac{1}{2},2)} \sigma^z_{(0,-\frac{1}{2},2)} \sigma^z_{(0,0,\frac{5}{2})} \sigma^z_{(0,0,\frac{1}{2})}$, that is not connected, without such modifications. Similarly, we can freely add $B_{\gamma_d}$ terms with $x_D = \frac{3}{2}$ to $H_1$ as such terms can be directly obtained by taking the products of $B_{\gamma_d}$ terms with $x_D = \frac{1}{2}$ and corresponding $\sigma^z_i \sigma^z_{i+I_D}$ terms. Besides, $H_{dddD}$ contains no $B_{\gamma_d}$ terms on the $T^{D-1}$. After that, we can see that all terms in $H_1$ still commute with each other. From another perspective, $H_1$ can also be be obtained by considering the conjugate action of the $\mathtt{CNOT}$ gates applied in the first step in Sec.~\ref{subsec:dddD_ER}. A more detailed demonstration of the terms in $H_1$ is given in Appendix~\ref{app_proof}.
 
Secondly, because the $\mathcal{U}$ transformation is a product of a series of $\mathtt{CNOT}$ gates, the conjugate action of $\mathcal{U}$ on an arbitrary stabilizer $G$ can be reduced to the conjugate action of $\mathtt{CNOT}$ gates on $G$. With the general mapping rules given by the conjugate action of $\mathtt{CNOT}$ gates (see Appendix~\ref{app_cnot}), we can obtain all terms in $H_2$ as follows:

\begin{enumerate}
	\item[$-$] First, since for an arbitrary $\gamma_{D-1}$ inside the $T^{D-1}$, all qubits that are controlled by the qubits from the $\gamma_{D-1}$ together with the control qubits themselves form a $\gamma_D$ with $x_D=\frac{1}{2}$, $A_{\gamma_{D-1}}$ terms in $H_{dddD-1}$ are mapped to $A_{\gamma_D}$ terms with $x_D=\frac{1}{2}$.
	\item[$-$] Second, since an arbitrary target qubit $\sigma_i$ with $x_D=\frac{1}{2}$ is controlled by exactly $2$ qubits from the $T^{D-1}$, each $\sigma^z_i \sigma^z_{i+I_D}$ term in $H_{zz}$ is mapped to a $4$-spin term composed of the original $\sigma^z_i$, $\sigma^z_{i+I_D}$ and the $z$ components of the $2$ qubits that control $\sigma_i$.
	\item[$-$] Third, further considering that an arbitrary target qubit $\sigma_i$ with $x_D=0$ is only controlled by $\sigma_{i+I_D}$ from the $T^{D-1}$, a $B_{\gamma_d}$ term in $H_{dddD}$ near the $T^{D-1}$ should be modified as follows: (a) for a qubit $\sigma_i$ with $x_D=0$ involved in the $B$ term, multiply the term by $\sigma^z_{i+I_D}$; (b) for a qubit $\sigma_i$ with $x_D=\frac{1}{2}$ involved in the $B$ term, multiply the term by the $z$ components of the $2$ qubits that control the $\sigma_i$. As an example, for a $B_{\gamma_d}$ term only involving qubits with $x_D=0$, it is mapped to a $B_{\gamma_d}B_{\gamma_d + I_D}$ term, where $B_{\gamma_d + I_D}$ is obtained by adding $I_D$ to the coordinates of all qubits involved in $B_{\gamma_d}$.
	\item[$-$] Finally, all other terms stay invariant under the conjugate action of $\mathcal{U}$.
\end{enumerate}

We denote the Hamiltonian of $[d,d+1,d+2,D]$ model on the lattice of the size $L_1\times L_2 \times \cdots \times (L_D + 1)$ with PBC as $H_3$ (see Eq.~(\ref{eq:dddD})). Then we can notice that by taking the product of $A$ terms obtained in the first step and $A'$ terms from $H_{dddD}$ we can obtain all $A$ terms that exist in $H_3$ but superficially missing in $H_2$; by taking the product of the $4$-spin terms obtained in the second step (which can be recognized as $B$ terms in $H_3$), modified $B$ terms obtained in the third step and $B$ terms in $H_{dddD-1}$ we can obtain all $B$ terms that exist in $H_3$ but superficially missing in $H_2$. Therefore, all terms of $H_3$ can be obtained by taking the product of terms of $H_2$ (a more detailed demonstration is given in Appendix~\ref{app_proof}). As it is also straightforward to check the other way around, finally, we can see that $H_2$ and $H_3$ are equivalent Pauli stabilizer code models with equivalent stabilizer groups, and $|\xi_f\rangle = \mathcal{U}|\xi_2\rangle$ is indeed a ground state of $H_3$. 

\subsection{Discussions}
\label{subsec:dis}

As we have demonstrated in this section, in the ERG transformations of different $[d,d+1,d+2,D]$ states, the added/removed states are also different LRE states. From another word, the entanglement patterns in $[d,d+1,d+2,D]$ states with different $D$ and a fixed $d$ are intrinsically different, and these models cannot be fully understood as fixed points of a finite number of types of ERG transformations. Instead, we need an infinite series of ERG transformations of different levels to understand the more general long range entanglement patterns.

Therefore, we conclude the above observations by proposing the concept of a hierarchy of ERG transformations, where each transformation is assigned with an integer level. Correspondingly, LRE states are assigned with integer levels as well (see Fig.~\ref{fig:tree}). For a given stabilizer code model considered in this paper, two level-$(n+1)$ LRE ($\mathtt{LRE}^{n+1}$) states of different sizes can be connected by a level-$n$ ERG ($\mathtt{ERG}^{n}$) transformation, that is composed of LU transformations combined with addition/removal of level-$n$ LRE states. Furthermore, if we define product states and short range entangled states  as $\mathtt{LRE}^0$ states, then we have $[0,1,2,2]$ (2D toric code) states as $\mathtt{LRE}^1$ states, $[0,1,2,3]$ (3D X-cube) states as $\mathtt{LRE}^2$ states, $[0,1,2,4]$ states as $\mathtt{LRE}^3$ states and so on. Besides, we can see that low level LRE states themselves can be recognized as trivial high level LRE states, just like a product state is recognized as a trivial ``pure'' topological order. Specially, a decoupled stack of $\mathtt{LRE}^n$ states is also a trivial $\mathtt{LRE}^{n+1}$ state, as it can reduced to nothing under a $\mathtt{ERG}^n$ transformation.

\section{Summary and outlook}
\label{sec:outlook}

In this paper,  by considering a class of Pauli stabilizer codes, 
 we constructed a more unified ERG framework  through    adding/removing  more general degrees of freedom. The well-established ERG processes of   the $[0,1,2,2]$ (2D toric code) and $[0,1,2,3]$ (3D X-cube) model are naturally included  as the simplest cases.   All Pauli stabilizer codes considered here are categorized into  a series of ``state towers'' as shown in Fig.~\ref{fig:tree};   in each tower, lower LRE states of level-$n$ are   added/removed in the level-$n$ ERG process of an upper LRE state of level-$(n+1)$. Several future directions are listed below.
 
First, we may expect a more general ERG framework shown in Eq.~(\ref{eq_ERG_relation_general}) can be constructed in other stabilizer codes.

Second, the completeness of the concept of level of LRE states   needs further exploration. For example, for type-II fracton ordered states~\cite{Haah2011,Vijay2016,Vijay2015}, such as Haah's code~\cite{Haah2011}, a series of ERG transformations have been constructed and studied~\cite{Dua2020,Haah2014,Swingle2016}, nevertheless, whether it is possible to consistently assign a level to such Type-II fracton ordered states and corresponding ERG transformations is yet to be determined.  Some further discussion about the hierarchy of ERG transformations may be beneficial for a more complete understanding of the entanglement patterns in more generic fracton orders.

Third, except for the stabilizer code models considered in this paper, physically, we can also consider models perturbed by external fields, which are no longer exactly solvable. Constructing ERG transformations for such models to investigate their fixed points is also an interesting direction. And some numerical techniques may also be useful in the study of such models~\cite{Muehlhauser2020,2022PhRvR...4c3111Z,Zhu2022}.

Finally, it is known that the ERG transformations are related to   MERA, that is a kind of tensor networks capable of efficiently encoding the entanglement signatures of certain quantum many-body states~\cite{Vidal2008,Koenig2009,Evenbly2014,Evenbly2014a}. In Ref.~\cite{Shirley2018}, it has been noticed that $[0,1,2,3]$ (3D X-cube) states bear exact branching MERA representations. Then it is natural to ask whether LRE states of general levels can have such tensor network representations. If so, the holographic geometries generated by such tensor networks are also worth exploring~\cite{Evenbly2011,Swingle2012a,Evenbly2017}.

\acknowledgements
This work was supported by NSFC Grant  No.~12074438, Guangdong Basic and Applied Basic Research Foundation under Grant No.~2020B1515120100, the Fundamental Research Funds for Central Universities (22qntd3005), and the Open Project of Guangdong   Provincial Key Laboratory of Magnetoelectric Physics and Devices under Grant No.~2022B1212010008.

%\bibliography{tem}

\appendix

\section{A brief introduction of controlled-NOT ($\mathtt{CNOT}$) gate}

\label{app_cnot}

Here we give a brief introduction of the controlled-NOT ($\mathtt{CNOT}$) gate that is to be frequently used in the main text of this paper.

By definition, $\mathtt{CNOT}$ gate is a 2-qubit unitary operation. In $\sigma^z$ basis, for $|x\ket, |y\ket$, where $x,y \in\{0,1\}$, $\mathtt{CNOT}$ gate maps $|x\ket \Otimes |y\ket$ to $|x\ket \Otimes |y\oplus x\ket$, here $\Otimes$ means tensor product, $\oplus$ means modulo $2$ addition. Effectively, $\mathtt{CNOT}$ gate regards the first qubit as a \textit{control qubit}, and the second qubit as a \textit{target qubit}. When the control (first) qubit is $|0\ket$, then $\mathtt{CNOT}$ gate does nothing; when the control qubit is $|1\ket$, $\mathtt{CNOT}$ gate flips the target (second) qubit, thus the name. For example, denoting the action of $\mathtt{CNOT}$ gate as $U$, we have $U|01\rangle=|01\rangle$ and $U|11\rangle=|10\rangle$. 

\begin{comment}
The unitarity of $\mathtt{CNOT}$ gate can be proved as follows: firstly, it is obvious that $U_{\text{CNOT}} U_{\text{CNOT}} = I$ ($I$ is the identity operator), thus $U_{\text{CNOT}} = U_{\text{CNOT}}^{-1}$. Then, as a $\mathtt{CNOT}$ gate is only supported on two qubits, we focus on a $2$-qubit Hilbert space. For an arbitrary pair of Ising configuration basis of this Hilbert space $|\psi_1\ket$ and $|\psi_2\ket$, we can check that $\bra \psi_1| U_{\text{CNOT}} |\psi_2\ket = \bra \psi_2| U_{\text{CNOT}} |\psi_1\ket^*$ (superscript $*$ is for complex conjugate): we only have $\bra \psi_1| U_{\text{CNOT}} |\psi_2\ket = 1$ when the control qubits in $|\psi_1\ket$ and $|\psi_2\ket$ are the same, and only the qubits to be flipped\footnote{Here ``qubits to be flipped'' means the target qubits when the control qubits are in $|1\ket$ states, and no qubit otherwise.} are different in $|\psi_1\ket$ and $|\psi_2\ket$; otherwise, $\bra \psi_1| U_{\text{CNOT}} |\psi_2\ket = 0$. That is to say, we have $U^{\dagger}_{\text{CNOT}} = U_{\text{CNOT}}$. As a result, we see that $\mathtt{CNOT}$ gate is both Hermitian and unitary.
\end{comment}

For the usage in the main text, here we also introduce the conjugate action of $\CNOT$ gate on stabilizers (i.e., Hamiltonian terms of a stabilizer code model and their products). For a state $|\phi\ket$ in the stabilizer subspace and a stabilizer $G$ (i.e. $G|\phi\ket=|\phi\ket$), if we apply a $\mathtt{CNOT}$ gate $U$ on $|\phi\ket$, then we have $(UGU^{\dagger}) U|\phi\ket=U|\phi\ket$. That is to say, the transformed state $U|\phi\ket$ is stabilized by $UGU^{\dagger}$, that is $G$ acted by the conjugate action of the $\mathtt{CNOT}$ gate. For a specific $G$ acting non-trivially on some control or target qubits, the correspondence between $G$ and $UGU^{\dagger}$ can be expressed as follows\cite{Shirley2018,Aguado2008}:
\begin{align*}
	ZI&\rightarrow ZI \\
	IZ&\leftrightarrow ZZ\\
	XI&\leftrightarrow XX\\
	IX&\rightarrow IX, 
\end{align*}
where the first qubit refers to the control qubit, and the second qubit refers to the target qubit. For example, if we consider the conjugate action of a $\mathtt{CNOT}$ gate on a stabilizer $G$, where $G$ applies a $\sigma^x$ on the control qubit, and applies an identity on the target qubit, then the corresponding $UGU^{\dagger}$ will apply $\sigma^x$ on both qubits. 

\section{Proof of the equivalence between Hamiltonians $H_2$ and $H_3$}
\label{app_proof}
	
In this appendix, we concretely demonstrate that in Sec.~\ref{subsec:proof_ham}, all terms in $H_{3}$, the Hamiltonian of $[d,d+1,d+2,D]$ model on
the lattice of the size $L_{1}\times L_{2}\times\cdots\times(L_{D}+1)$
with PBC, can be obtained by taking the product of terms in $H_{2}$ and vice versa, thus they are equivalent stabilizer code models.
As we can notice that $H_{2}$ and $H_{3}$ only have different terms
around the $T^{D-1}$ with $x_{D}=1$, we only need to consider
terms defined on locations with $0\leq x_{D}\leq2$.  

Before discussing about terms in $H_{2}$, we would to like to give a detailed demonstration and classification of the terms around the cut $T^{D-1}$ in $H_{1}=H_{dddD}+H_{dddD-1}+H_{zz}$. Such terms
in $H_{1}$ can be classified as follows:
\begin{itemize}
	\item $BI$ terms: $B$ terms with $x_{D}=0$ from $H_{dddD}$ that only
	involve qubits with $x_{D}=0$.
	\item $BII$ terms: $B$ terms with $x_{D}=0$ from $H_{dddD}$ that simultaneously
	involve qubits with $x_{D}=0$ and $x_{D}=\frac{1}{2}$.
	\item $BIII$ terms: $B$ terms with $x_{D}=\frac{1}{2}$ from $H_{dddD}$.
	\item $BIV$ terms: $B$ terms with $x_{D}=1$ from $H_{dddD-1}$ (i.e.
	such $B$ terms only involve qubits with $x_{D}=1$).
	\item $BV$ terms: $B$ terms with $x_{D}=\frac{3}{2},2$ from $H_{dddD}$.
	\item $AI$ terms: $A'_{\gamma_{D}}=A{}_{\gamma_{D}}A{}_{\gamma_{D}+I_{D}}$
	terms with $x_{D}=\frac{1}{2}$ from $H_{dddD}$.
	\item $AII$ terms: $A_{\gamma_{D-1}}$terms with $x_{D}=1$ from $H_{dddD-1}$.
	\item $C$ terms: $C_{\gamma_{d+1}}=\sigma_{\gamma_{d+1}}^{z}\sigma_{\gamma_{d+1}+I_{D}}^{z}$
	with $x_{D}=\frac{1}{2}$ from $H_{zz}$.
\end{itemize}

And we can notice that, around the $T^{D-1}$, $H_{3}$ is composed of $BI$, $BII$, $BIII$, $BIV$, $BV$, and the following
terms:
\begin{itemize}
	\item $BVI$ terms: $B$ terms with $x_{D}=1$ that involve qubits with
	$x_{D}\neq1$.
	\item $AIII$ terms: $A_{\gamma_{D}}$ terms with $x_{D}=\frac{1}{2},\frac{3}{2}$. 
\end{itemize}

Then, we consider the conjugate action of the LU transformation $\mathcal{U}$ on the terms in $H_{1}$, that leads to terms in $H_2$ (note that here the superscripts of $B$ terms are omitted as we only need to consider the types of terms):
\begin{itemize}
	\item A $BI$ term $B_{\gamma_{d}}$ is mapped to $B_{\gamma_{d}}B_{\gamma_{d}+I_{D}}$,
	the product of the $BI$ term itself and a $BIV$ term $B_{\gamma_{d}+I_{D}}$.
	\item A $BII$ term $B_{\gamma_{d}}$ is mapped to (a) the $BII$ term
	itself, if the qubits with $x_{D}=0$ and $x_{D}=\frac{1}{2}$ in
	$B_{\gamma_{d}}$ are controlled by the same pair of qubits from the $T^{D-1}$; (b) $B_{\gamma_{d}}B_{\gamma_{d}+I_{D}}$, the
	product of the $BII$ term itself and a $BIV$ term $B_{\gamma_{d}+I_{D}}$,
	if otherwise.
	\item A $BIII$ term $B_{\gamma_{d}}$ is mapped to (a) the $BIII$ term itself, if $2$ perpendicular qubits in $B_{\gamma_{d}}$ (i.e., the
	$2$ qubits are nearest and from different $(d+1)$-dimensional subsystems)
	share $1$ control qubit; (b) $B_{\gamma_{d}}B_{\gamma_{d}^{1}}B_{\gamma_{d}^{2}}$,
	the product of the $BIII$ term itself and $2$ $BIV$ terms $B_{\gamma_{d}^{1}}$
	and $B_{\gamma_{d}^{2}}$, if $2$ perpendicular qubits in $B_{\gamma_{d}}$
	have control qubits that nearest to the same $\gamma_{d}$; (c) $B_{\gamma_{d}}B_{\gamma_{d}^{1}}B_{\gamma_{d}^{2}}B_{\gamma_{d}^{3}}B_{\gamma_{d}^{4}}$,
	the product of the $BIII$ term itself and $4$ $BIV$ terms $B_{\gamma_{d}^{1}}$,
	$B_{\gamma_{d}^{2}}$, $B_{\gamma_{d}^{3}}$ and $B_{\gamma_{d}^{4}}$,
	if otherwise;
	\item $BIV$, $BV$, $AI$ terms stay invariant.
	\item An $AII$ term $A_{\gamma_{D-1}}$terms with $x_{D}=1$ is mapped
	to an $AIII$ term $A_{\gamma_{D}}$ with $x_{D}=\frac{1}{2}$, where
	the $\gamma_{D}$ is obtained by $\gamma_{D}=\gamma_{D-1}-\frac{1}{2}I_{D}$.
	\item A $C$ term $C_{\gamma_{d+1}}$ term is mapped to a $BVI$ term $B_{\gamma_{d}}$
	with $x_{D}=1$, where the $\gamma_{d}$ is obtained as $\gamma_{d}=\gamma_{d+1}+\frac{1}{2}I_{D}$.
\end{itemize}
Because $BIV$, $BV$, $AI$ terms are invariant under the conjugate
action of $\mathcal{U}$ (i.e., they present in $H_{2}$), we can obtain an
arbitrary $BI$, $BII$ or $BIII$ term by taking the product of the
corresponding transformed term with invariant $BIV$ terms. An arbitrary $AIII$ term with $x_{D}=\frac{1}{2}$ can be obtained as a transformed $AII$ term, and an arbitrary $AIII$ term with $x_{D}=\frac{3}{2}$ can be obtained by
taking the product of a transformed $AII$ term and an invariant $AI$
term. Besides, by taking the product of a transformed $C$ term with
a $BIV$ term, an arbitrary $BVI$ term can also be obtained. Finally,
because it is straightforward to check that all terms in $H_{2}$ can be
obtained by taking products of terms in $H_{3}$, as two stabilizer
code models, $H_{2}$ and $H_{3}$ have equivalent sets of stabilizer generators, thus the stabilizer subspaces should be equivalent. 

\section{Another level-$0$ ERG transformation of $[0,1,2,2]$ states}\label{appendix_toriccode}

\begin{figure}
	\centering
	\includegraphics[width=0.8\linewidth]{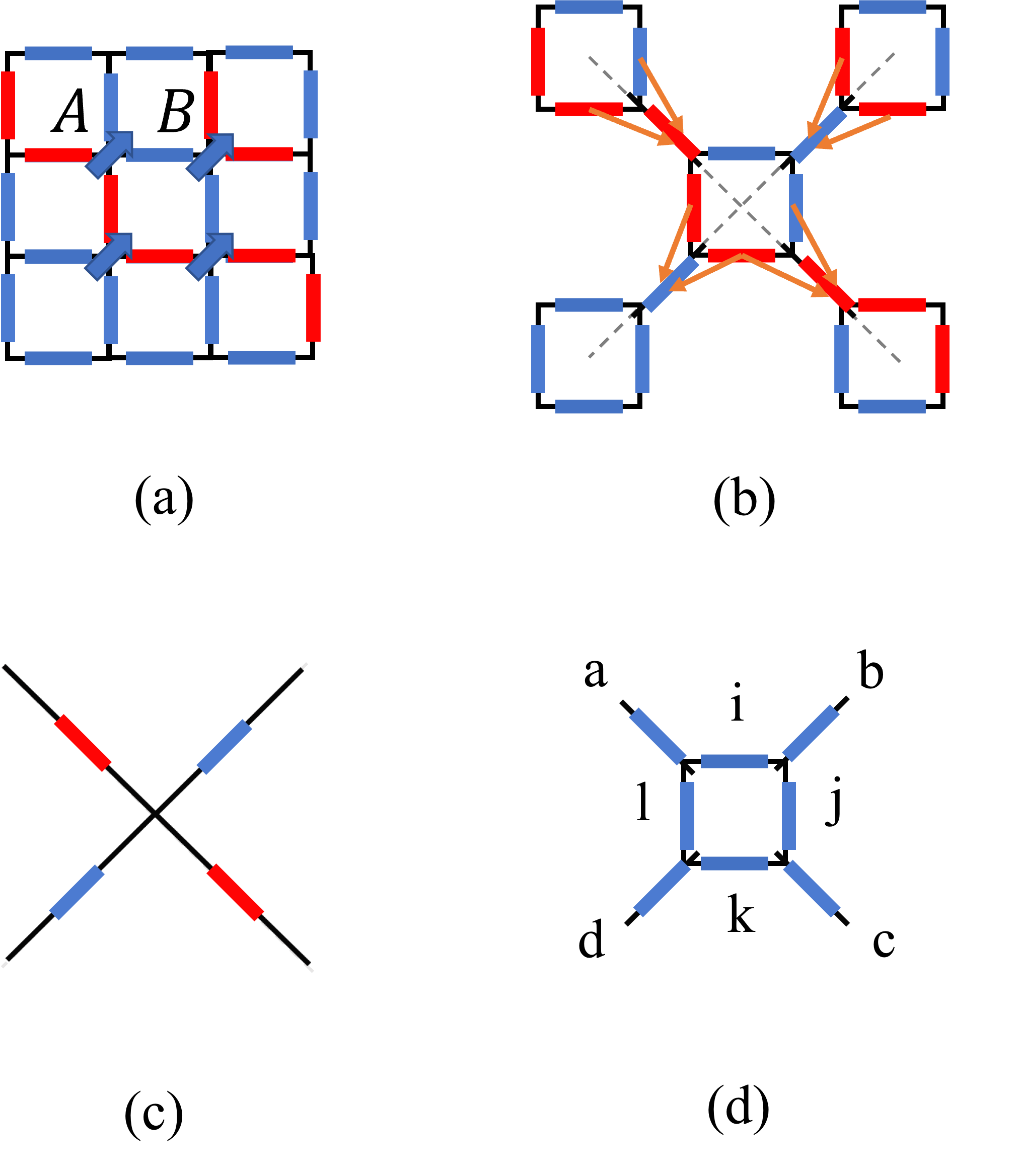}
	\caption{\textbf{Another ERG transformation of the 2D toric code model labeled by $[0,1,2,2]$.}    This ERG transformation is denoted as $\mathtt{ERG}^0$ in Eq.~(\ref{eq_ERG_relation_ToricCode}).  In (a), we demonstrate how the original vertices are separated into two sublattices.  A closed string configuration is illustrated, where $|0\ket$ spins on the links are denoted by blue bars, and $|1\ket$ spins forming strings are highlighted with red. The four blue arrows on four vertices denote four additional spins in state $|0\ket$. In (b), we can see the four vertices are now extended to four links connecting plaquettes, and their corresponding additional spins, which are also denoted by bars now, \textit{have been} transformed by $\mathtt{CNOT}$ gates such that the strings (formed by $|1\ket$ spins) are still closed. These $\mathtt{CNOT}$ gates targeting on the four additional spins are denoted by orange arrows pointing from control qubits to target qubits. Besides, dashed lines connecting the centers of squares are presented. As we can see, the action of $\mathtt{CNOT}$ gates couples additional and original spins in a manner that  indeed preserves the closed strings pattern of $[0,1,2,2]$ (2D toric code) state. By dropping all spins nearest to squares after another LU transformation, we  obtain (c) in which a square lattice with a larger lattice constant appears and spins are located at the centers of new links (i.e., the dashed lines in (b)). We can see that, now spins on the centers of new links form a $[0,1,2,2]$ state on a new square lattice. In (d) we demonstrate an assignment of labels to the eight links around a square, where diagonal links are denoted by $a$, $b$, $c$ and $d$, links nearest to the square are denoted by $i$, $j$, $k$ and $l$.}
	\label{fig_toriccode}
\end{figure}

In this appendix, we review the $\ERG^0$ transformation of $[0,1,2,2]$ (2D toric code) states following the recipe in Ref.~\cite{Zeng2019}. In this $\mathtt{ERG}^0$ transformation of a $[0,1,2,2]$ state defined on a square lattice with PBC, we firstly separate vertices into A and B sublattices. Then, we put an additional $1/2$-spin in state $|0\rangle$ on each vertex  (see Fig.~\ref{fig_toriccode}(a)). After that, we apply an LU transformation $\mathcal{U}_1$, that is composed of a series of $\mathtt{CNOT}$ gates: for each additional spin, we act two $\mathtt{CNOT}$ gates targeting on it. More specifically, for an additional qubit in sublattice A, we use the qubits on the up and left links as control qubits; for an additional qubit in sublattice B, we use qubits on the up and right links  (see Fig.~\ref{fig_toriccode}(b)). The action of $\mathcal{U}_1$ can be understood pictorially: recall that in any allowed Ising configuration of a $[0,1,2,2]$ state there must be either an even number or zero of $|1\ket$ spins around each vertex, thus flipped spins always form closed strings. Then, we can notice that the design of $\mathtt{CNOT}$ gates in $\mathcal{U}_1$ exactly preserves this constraint by integrating additional spins into the closed strings pattern. 

After that, we further apply an LU transformation $\mathcal{U}_2$ to map the spins around each square to $|0000\ket + |1111\ket$ (normalization is omitted), and such spins can be removed out of the state as $|0000\ket + |1111\ket$ can be transformed to a product state by a local unitary operator. For the lattice, the LU transformation $\mathcal{U}_2$ and the removal of spins effectively shrinks every square to a vertex as shown in Fig.~\ref{fig_toriccode}(c). By noticing that in each configuration there is always an even number or zero of diagonal links around each square with qubits in $|1\ket$, the resulting state also has the closed strings pattern. Here, to see that $\mathcal{U}_2$ is indeed an LU transformation, we can recognize $\mathcal{U}_2=\prod_s \mathcal{U}_s$, which is the product of $\mathcal{U}_s$ operators supported around each square $s$. A $\mathcal{U}_s$ acts on the four qubits on links nearest to the square $s$ (denoted by $i$, $j$, $k$ and $l$) and the four qubits on the diagonal links around $s$ (denoted by $a$, $b$, $c$ and $d$, see Fig.~\ref{fig_toriccode}(d)). Here, $\mathcal{U}_s$ can  be roughly recognized as a generalized $\mathtt{CNOT}$ gate: it takes the qubits on diagonal links as control qubits, and qubits on the square as targets. For a specific configuration of the eight qubits, the action of $\mathcal{U}_s$ can be obtained as follows: a) if all control qubits are $|0\ket$, then flip no target qubits; b) if two control qubits are $|1\ket$, then flip the target qubits between them clockwise following the alphabetical order (e.g. if qubits on $b$ and $d$ are $|1\ket$, then flip qubits on $j$ and $k$); c) if all control qubits are in $|1\ket$, then flip qubits on $i$ and $k$. Then, $\mathcal{U}_s$ obviously satisfies $\mathcal{U}_s \mathcal{U}_s = \mathbb{I}$, thus $\mathcal{U}^{-1}_s = \mathcal{U}_s$. Next, notice that Ising configurations form a complete basis of the Hilbert space, for an arbitrary pair of Ising configurations of the eight qubits $|\psi_1\ket$ and $|\psi_2\ket$, we can obtain that $\bra \psi_1 | \mathcal{U}_s|\psi_2\ket = \bra \psi_2 | \mathcal{U}_s  |\psi_1\ket^*$: we only have $\bra \psi_1 | \mathcal{U}_s |\psi_2\ket=1$ when the control qubits in $|\psi_1\ket$ and $|\psi_2\ket$ are all the same, and only the qubits to be flipped are different in  $|\psi_1\ket$ and $|\psi_2\ket$; otherwise, $\bra \psi_1 | \mathcal{U}_s |\psi_2\ket=0$. As a result, $\mathcal{U}_s^{\dagger} = \mathcal{U}_s = \mathcal{U}_s^{-1}$, thus $\mathcal{U}_s$ is both unitary and Hermitian. Since the transformations above do not change the pattern that the state is invariant under the action of $A_p$ terms on squares, and $\mathcal{U}_s$ always maps the configuration of a square to $|0000\ket$ or $|1111\ket$, we can see that spins nearest to each square are indeed mapped to $|0000\ket + |1111\ket$. 

Finally, we obtain a $[0,1,2,2]$ state on a square lattice with a larger lattice constant. That is to say, after an ERG transformation composed of adding/removing product states and LU transformations, the structure of the $[0,1,2,2]$ state is preserved. Or from another perspective, the $[0,1,2,2]$ model is a fixed point of the $\ERG^0$ transformation as symbolically expressed in Eq.~(\ref{eq_ERG_relation_ToricCode}). A pictorial demonstration of this ERG transformation is given in Fig.~\ref{fig_toriccode}.

%\bibliography{tem}

\end{document}